\documentstyle[12pt,epsf]{article}
\def\dps{\displaystyle}
\def\mib#1{\mbox{\boldmath $#1$}}
\def\sla#1{\mbox{$#1\!\!\scriptstyle{/}$}}
\def\slaq{\mbox{$q\!\!\!\scriptstyle{/}$}\,}
\def\bra#1{\langle #1 |} \def\ket#1{|#1\rangle}
\def\vev#1{\langle #1\rangle}
\def\sst#1{\scriptscriptstyle{#1}}
\def\ssf{\scriptscriptstyle{f}}
\def\ssl{\scriptscriptstyle{\ell}}
\def\ssb{\scriptscriptstyle{b}}
\def\xf{x_{\!\scriptscriptstyle{f}}}
\def\thf{\theta_{\!\scriptscriptstyle{f}}}
\newcommand{\nc}{\newcommand}
\nc{\postscript}[2] 
{\setlength{\epsfxsize}{#2\hsize}\centerline{\epsfbox{#1}}}
\nc{\bg}{B. Grzadkowski}
\nc{\non}{\nonumber}
\nc{\barx}{\bar{x}}\nc{\pbarn}{\;\hbox {pb}}\nc{\fbarn}{\;\hbox {fb}}
\nc{\hc}{\hbox {h.c.}} 
\nc{\re}{\hbox {Re}} 
\nc{\im}{\hbox {Im}}
\nc{\mev}{\hbox {MeV}} \nc{\gev}{\;\hbox {GeV}} 
\nc{\tev}{\;\hbox {TeV}}
\def\gesim{\lower0.5ex\hbox{$\:\buildrel >\over{\scriptstyle\sim}\:$}} 
\def\lesim{\lower0.5ex\hbox{$\:\buildrel <\over{\scriptstyle\sim}\:$}} 
\nc{\prd}[3]{{\it Phys.\ Rev.}\ {{\bf D{#1}} (#2), #3}}
\nc{\prl}[3]{{\it Phys.\ Rev.\ Lett.}\ {{\bf {#1}} (#2), #3}}
\nc{\plb}[3]{{\it Phys.\ Lett.}\ {{\bf B{#1}} (#2), #3}}
\nc{\npb}[3]{{\it Nucl.\ Phys.}\ {{\bf B{#1}} (#2), #3}}
\nc{\ptp}[3]{{\it Prog.\ Theor.\ Phys.}\ {{\bf {#1}} (#2), #3}}
\nc{\zfp}[3]{{\it Z.\ Phys.}\ {{\bf C{#1}} (#2), #3}}
\nc{\mpla}[3]{{\it Mod.\ Phys.\ Lett.}\ {{\bf A{#1}} (#2), #3}}
\nc{\rmp}[3]{{\it Rev.\ Mod.\ Phys.}\ {{\bf {#1}} (#2), #3}}
\nc{\ijmpa}[3]{{\it Int.\ J.\ Mod.\ Phys.}\
               {{\bf A{#1}} (#2), #3}}
\nc{\epj}[3]{{\it Eur.\ Phys.\ J.}\ {{\bf C{#1}}  (#2), #3}}

\nc{\ttbar}{t\bar{t}}         \nc{\bbbar}{b\bar{b}}
\nc{\tanb}{\tan \beta}        \nc{\twbdec}{t\to W^+ b}
\nc{\tbwbdec}{\bar{t}\to W^- \bar{b}}
\nc{\epem}{e^+e^-}            \nc{\eett}{\epem \to \ttbar}
\nc{\sigeett}{\sigma_{e\bar{e}\to\ttbar}}
\nc{\wpwm}{W^+W^-}            \nc{\tbar}{\bar{t}}
\nc{\bbar}{\bar{b}}           \nc{\wpp}{W^+}
\nc{\mt}{m_t}    \nc{\mts}{m_t^2}   \nc{\mw}{M_W}    \nc{\mws}{M_W^2}
\nc{\mz}{M_Z}    \nc{\mzs}{M_Z^2}
\nc{\ttbardec}{\ttbar \to W^+W^-\bbbar}
\nc{\wwbb}{W^+W^-\bbbar}      \nc{\sm}{SM}
\nc{\cw}{\cos\theta_W}        \nc{\sw}{\sin\theta_W}
\nc{\sws}{\sin^2\theta_W}     \nc{\sig}{\sigma_{tot}}
\nc{\lp}{{\ell}^+}              
\nc{\lm}{{\ell}^-}
\nc{\lpm}{{\ell}^\pm}
\nc{\tb}{\stackrel{{\scriptscriptstyle (-)}}{t}}
\nc{\bb}{\stackrel{{\scriptscriptstyle (-)}}{b}}
\nc{\fb}{\stackrel{{\scriptscriptstyle (-)}}{f}}
\nc{\epsl}{\epsilon_L}        \nc{\cp}{C\!P}

\nc{\splus}{s_+}       \nc{\smin}{s_-}        \nc{\eps}{\epsilon}
\nc{\psp}{Ps_+}        \nc{\psm}{Ps_-}        \nc{\lsp}{ls_+}
\nc{\lsm}{ls_-}        \nc{\sss}{s_+s_-}      \nc{\m}{m_t}
\nc{\mq}{m_t^2}        \nc{\mr}{\frac{1}{\m}} \nc{\av}{A_{\gamma}}
\nc{\bv}{B_{\gamma}}   \nc{\az}{A_Z}          \nc{\bz}{B_Z}
\nc{\avs}{A_{\gamma}^2}\nc{\azs}{A_Z^2}       \nc{\bzs}{B_Z^2}
\nc{\dav}{\delta \! A_{\gamma}}   \nc{\dbv}{\delta \! B_{\gamma}}
\nc{\dcv}{\delta C_{\gamma}}      \nc{\ddv}{\delta \! D_{\gamma}}
\nc{\daz}{\delta \! A_Z}          \nc{\dbz}{\delta \! B_Z}
\nc{\dcz}{\delta C_Z}             \nc{\ddz}{\delta \! D_Z}
\nc{\dev}{\delta \! E_{\gamma}}   \nc{\dez}{\delta \! E_Z}
\nc{\dfv}{\delta \! F_{\gamma}}   \nc{\dfz}{\delta \! F_Z}
\nc{\rdav}{{\rm Re}(\delta \! A_{\gamma}) \:}
\nc{\rdbv}{{\rm Re}(\delta \! B_{\gamma}) \:}
\nc{\rdcv}{{\rm Re}(\delta C_{\gamma}) \:}
\nc{\rddv}{{\rm Re}(\delta \! D_{\gamma}) \:}
\nc{\rdaz}{{\rm Re}(\delta \! A_Z) \:}
\nc{\rdbz}{{\rm Re}(\delta \! B_Z) \:}
\nc{\rdcz}{{\rm Re}(\delta C_Z) \:}
\nc{\rddz}{{\rm Re}(\delta \! D_Z) \:}
\nc{\idav}{{\rm Im}(\delta \! A_{\gamma}) \:}
\nc{\idbv}{{\rm Im}(\delta \! B_{\gamma}) \:}
\nc{\idcv}{{\rm Im}(\delta C_{\gamma}) \:}
\nc{\iddv}{{\rm Im}(\delta \! D_{\gamma}) \:}
\nc{\idaz}{{\rm Im}(\delta \! A_Z) \:}
\nc{\idbz}{{\rm Im}(\delta \! B_Z) \:}
\nc{\idcz}{{\rm Im}(\delta C_Z) \:}
\nc{\iddz}{{\rm Im}(\delta \! D_Z) \:}
\nc{\cz}{(1+v_e^2)d\:\!'^2}         \nc{\ci}{v_ed\:\!'}
\nc{\ccz}{v_ed\:\!'^2}              \nc{\cci}{d\:\!'}
\nc{\dxdcos}{{d^2\sigma \over d\xf\;d\cos\thf}}
\nc{\gl}{{\mit\Gamma}_{\ell}}         \nc{\gw}{{\mit\Gamma}_W}
\nc{\gf}{{\mit\Gamma}_{\sst{f}}}      \nc{\gb}{{\mit\Gamma}_b} 
\nc{\reaf}{\re(f_2^R)}
\nc{\bet}{\beta}                \nc{\bs}{\bet^2}
\nc{\thetaf}{\theta_f}
\nc{\lspace}{\;\;\;\;\;\;\;\;\;\;}  \nc{\llspace}{\lspace \lspace}
\nc{\beq}{\begin{equation}}   \nc{\eeq}{\end{equation}}
\nc{\bea}{\begin{eqnarray}}   \nc{\eea}{\end{eqnarray}}
\nc{\baa}{\begin{array}}      \nc{\eaa}{\end{array}}
\nc{\bit}{\begin{itemize}}    \nc{\eit}{\end{itemize}}
\nc{\ben}{\begin{enumerate}}  \nc{\een}{\end{enumerate}}
\nc{\bce}{\begin{center}}     \nc{\ece}{\end{center}}
\begin{document}
\pagestyle{empty} \setlength{\footskip}{2.0cm}
\setlength{\oddsidemargin}{0.5cm} \setlength{\evensidemargin}{0.5cm}
\renewcommand{\thepage}{-- \arabic{page} --}
\def\mib#1{\mbox{\boldmath $#1$}}
\def\bra#1{\langle #1 |}      \def\ket#1{|#1\rangle}
\def\vev#1{\langle #1\rangle} \def\dps{\displaystyle}
\def\thebibliography#1{\centerline{REFERENCES}
     \list{[\arabic{enumi}]}{\settowidth\labelwidth{[#1]}\leftmargin
     \labelwidth\advance\leftmargin\labelsep\usecounter{enumi}}
     \def\newblock{\hskip .11em plus .33em minus -.07em}\sloppy
     \clubpenalty4000\widowpenalty4000\sfcode`\.=1000\relax}\let
     \endthebibliography=\endlist
\def\sec#1{\addtocounter{section}{1}\section*{\hspace*{-0.72cm}
     \normalsize\bf\arabic{section}.$\;$#1}\vspace*{-0.3cm}}
\renewcommand{\theequation}{\thesection.\arabic{equation}}
\vspace*{-1.5cm}
\begin{flushright}
$\vcenter{
\hbox{IFT-04-00}
\hbox{TOKUSHIMA Report} 
\hbox{(hep-ph/0004223)}
}$
\end{flushright}

\vskip 0.5cm
\begin{center}
{\large\bf Optimal-Observable Analysis of the Angular and}
\vskip 0.1cm
{\large\bf Energy Distributions for Top-Quark Decay Products}
\vskip 0.1cm
{\large\bf at Polarized Linear Colliders}
\end{center}

\vspace*{0.7cm}
\begin{center}
\renewcommand{\thefootnote}{\alph{footnote})}
{\sc Bohdan GRZADKOWSKI$^{\:1),\:}$}\footnote{E-mail address:
{\tt bohdan.grzadkowski@fuw.edu.pl}}\ and\ 
{\sc Zenr\=o HIOKI$^{\:2),\:}$}\footnote{E-mail address:
{\tt hioki@tokushima-u.ac.jp}}
\end{center}

\vspace*{0.7cm}
\centerline{\sl $1)$ Institute of Theoretical Physics,\ Warsaw 
University}
\centerline{\sl Ho\.za 69, PL-00-681 Warsaw, POLAND}

\vskip 0.3cm
\centerline{\sl $2)$ Institute of Theoretical Physics,\ 
University of Tokushima}
\centerline{\sl Tokushima 770-8502, JAPAN}

\vspace*{1.2cm}
\centerline{ABSTRACT}

\vspace*{0.4cm}
\baselineskip=20pt plus 0.1pt minus 0.1pt
An optimal-observable analysis of the angular and energy distributions
of the leptons and bottom quarks in the process $\epem \to \ttbar \to
{\ell}^\pm/\!\bb \cdots$ has been performed in order to measure the
most general top-quark couplings to gauge bosons at polarized linear
colliders. The optimal beam polarization for determination of each
coupling has been found. A very sensitive test of $C\!P$ violation in
$\ttbar$ production and decay has been proposed.
\\

\vspace*{0.7cm} 

\centerline{( Published in \npb{585}{2000}{3} )}

\vfill



\newpage
\renewcommand{\thefootnote}{\sharp\arabic{footnote}}
\pagestyle{plain} \setcounter{footnote}{0}
\baselineskip=21.0pt plus 0.2pt minus 0.1pt

\sec{Introduction}

In spite of the fact that the top quark has been discovered already 
several years ago \cite{top} its interactions are still very weakly
constrained. It remains an open question if top-quark couplings obey
the Standard Model (SM) scheme of the electroweak forces or there
exists a contribution from physics beyond the SM. We could interpret
the great success of the 1-loop precision tests of the SM as a strong
indication that the third generation also obeys the SM scheme.
However, an independent and direct measurement of the top-quark
couplings is definitely necessary before drawing any definite
conclusion concerning non-standard physics. 

Over the past several years there was a substantial effort devoted to
a possibility of determining top-quark couplings through measurements
performed at the open top region\footnote{Recently an interesting and
  complementary analysis by Jezabek, Nagano and Sumino has been
  published \cite{jez_CP} where the authors discussed possibility of
  determining $C\!P$-violating production form factors at the $\ttbar$
  threshold region.}\ 
of future $\epem$ linear colliders \cite{CKP}--\cite{Rest}. The
existing studies focused mainly on tests of $C\!P$ violation in
top-quark interactions.
In this article we will construct some new tools which could help to
measure both {\it $C\!P$ violating and $C\!P$ conserving} top-quark
couplings at linear colliders and therefore reveal the structure of
fundamental interactions beyond the SM. 

The top quark decays immediately after being produced and its huge
mass $m_t\simeq 174$ GeV leads to a decay width ${\mit\Gamma}_t$ much
larger than ${\mit\Lambda}_{\rm QCD}$. Therefore the decay process is
not influenced by any fragmentation effects \cite{Bigi} and decay
products will provide useful information on top-quark properties. Here
we will consider distributions of either ${\ell}^\pm$ in the inclusive
process $\epem \to \ttbar \to{\ell}^\pm \cdots$ or bottom quarks from
$\epem \to \ttbar \to \bb \cdots$. It turns out that the analysis of
the leptonic and $b$-quark final states is similar and could be
presented simultaneously. Although $\ttbar$ are also produced via $WW$
fusion \cite{Pil1}, we do not consider here this mechanism since
$\sigma(\epem \to \ttbar \nu \bar{\nu})$ is expected to be much smaller
than $\sigma(\epem \to \ttbar)$ for the energy of our interest 
($\sqrt{s} \lesim 2\tev$) \cite{nunu}.

This paper is organized as follows. First in sec.2 we describe the
basic framework of our analysis, and then show the angular and energy
distributions of the lepton and $b$-quark in sec.3. In sec.4, after
briefly reviewing the optimal-observable procedure \cite{opt}, we
estimate to what precision all the non-standard parameters can be
measured or constrained adjusting the initial beam polarizations.
Finally, we summarize our results in sec.5. In the appendix we collect
several functions used in the main text for completeness, though some
of them could also be found in our previous papers \cite{GH_npb,BGH}.

\sec{Framework and Formalism}
\label{frame} \setcounter{equation}{0}

We parameterize $\ttbar$ couplings to the photon and the $Z$ boson in
the following way
\begin{equation}
{\mit\Gamma}_{vt\bar{t}}^{\mu}=
\frac{g}{2}\,\bar{u}(p_t)\,\Bigl[\,\gamma^\mu \{A_v+\delta\!A_v
-(B_v+\delta\!B_v) \gamma_5 \}
+\frac{(p_t-p_{\bar{t}})^\mu}{2m_t}(\delta C_v-\delta\!D_v\gamma_5)
\,\Bigr]\,v(p_{\bar{t}}),\ \label{ff}
\label{prodff}
\end{equation}
where $g$ denotes the $SU(2)$ gauge coupling constant, $v=\gamma,Z$,
and 
\[
\av=\frac43\sw,\ \ \bv=0,\ \ 
\az=\frac1{2\cw}\Bigl(1-\frac83\sin^2\theta_W\Bigr),\ \ 
\bz=\frac1{2\cw}
\]
denote the SM contributions to the vertices. Among the above non-SM
form factors, $\delta\!A_v$, $\delta\!B_v$, $\delta C_v$ describe
$C\!P$-conserving while $\delta\!D_v$ parameterizes $C\!P$-violating
interactions. Similarly, we adopt the following parameterization of
the $Wtb$ vertex suitable for the $t$ and $\tbar$ decays:
\begin{eqnarray}
&&\!\!{\mit\Gamma}^{\mu}_{Wtb}=-{g\over\sqrt{2}}V_{tb}\:
\bar{u}(p_b)\biggl[\,\gamma^{\mu}(f_1^L P_L +f_1^R P_R)
-{{i\sigma^{\mu\nu}k_{\nu}}\over M_W}
(f_2^L P_L +f_2^R P_R)\,\biggr]u(p_t), \non \\
\label{dectff}
&&\!\!\bar{\mit\Gamma}^{\mu}_{Wtb}=-{g\over\sqrt{2}}V_{tb}^*\:
\bar{v}(p_{\bar{t}})
\biggl[\,\gamma^{\mu}(\bar{f}_1^L P_L +\bar{f}_1^R P_R)
-{{i\sigma^{\mu\nu}k_{\nu}}\over M_W}
(\bar{f}_2^L P_L +\bar{f}_2^R P_R)\,\biggr]v(p_{\bar{b}}),~~~~~
\label{dectbarff}
\end{eqnarray}
where $P_{L/R}=(1\mp\gamma_5)/2$, $V_{tb}$ is the $(tb)$ element of
the Kobayashi-Maskawa matrix and $k$ is the momentum of $W$. 
In the SM $f_1^L=\bar{f}_1^L=1$ and all the other form factors vanish.
On the
other hand, it is assumed here that interactions of leptons with gauge
bosons are properly described by the SM. Throughout the calculations
all fermions except the top quark are considered as massless. We also
neglect terms quadratic in the non-standard form factors.

Using the technique developed by Kawasaki, Shirafuji and Tsai
\cite{technique} one can derive the following formula for the
inclusive distributions of the top-quark decay product $f$ in the
process $\eett \to f + \cdots$~\cite{GH_npb}:
\begin{equation}
\frac{d^3\sigma}{d\mib{p}_{\ssf}/(2p_{\ssf}^0)}
(\epem \to f + \cdots)
=4\int d{\mit\Omega}_t
\frac{d\sigma}{d{\mit\Omega}_t}(n,0)\frac{1}{{\mit\Gamma}_t}
\frac{d^3{\mit\Gamma}_{\ssf}}{d\mib{p}_{\ssf}/(2p_{\ssf}^0)}
(t\to f + \cdots),
\label{master}
\end{equation}
where ${\mit\Gamma}_t$ is the total top-quark decay width and
$d^3{\mit\Gamma}_{\ssf}$ is the differential decay rate for the
process considered. $d\sigma(n,0)/d{\mit\Omega}_t$ is obtained from
the angular distribution of $\ttbar$ with spins $s_+$ and $s_-$ in
$\eett$, $d\sigma(s_+,s_-)/d{\mit\Omega}_t$, by the following
replacement:
\begin{equation}
s_{+\mu} \to n^{\ssf}_\mu=
-\Bigl[\:g_{\mu\nu}-\frac{{p_t}_\mu{p_t}_\nu}{m_t^2}\:\Bigr]
\frac{\sum\dps{\int} d{\mit\Phi}\:\bar{B}{\mit\Lambda}_+\gamma_5
\gamma^\nu B}{\sum\dps{\int}d{\mit\Phi}\:\bar{B}{\mit\Lambda}_+ B},
\ \ \ s_{-\mu}\to 0,
\label{n-vec}
\end{equation}
where the matrix element for $t(s_+)\to f+\cdots$ was expressed as
$\bar{B}u_t(p_t,s_+)$, ${\mit\Lambda}_+\equiv \sla{p}_t +m_t$,
$d{\mit\Phi}$ is the relevant final-state phase-space element and
$\sum$ denotes the appropriate spin summation.

\sec{Angular/Energy Distributions}
\label{distributions} \setcounter{equation}{0}

In this section we present $d^2\sigma/d\xf d\cos\thf$ for the
top-quark decay product $f(=\lpm/\!\bb)$, where $\xf$ denotes the
normalized energy of $f$ defined in terms of its energy $E_{\ssf}$
and the top-quark velocity $\beta(\equiv\sqrt{1-4m_t^2/s})$ as
$$
\xf \equiv \frac{2E_{\ssf}}{m_t}\sqrt{\frac{1-\beta}{1+\beta}}
$$
and $\thf$ is the angle between the $e^-$ beam direction and the $f$
momentum, all in the $\epem$ CM frame. 

Direct calculations performed in presence of the general decay vertex
(\ref{dectff}) lead to the following result for the $n^{\ssf}_\mu$
vector defined in eq.(\ref{n-vec}):
\begin{equation}
n^{\ssf}_\mu=
\alpha^{\ssf}\left(g_{\mu \nu}-\frac{p_{t \mu} p_{t \nu}}{\mts}
\right)\frac{\mt}{p_t p_{\ssf}}p_{\ssf}^\nu
\label{repl}
\end{equation}
where for a given final state $f$, $\alpha^{\ssf}$ is a calculable
depolarization factor 
\begin{equation}
\alpha^{\ssf}=\left\{
\begin{array}{ll}
1 &  {\rm for}\;\;f={\ell}^+  \\
  &                           \\
{\displaystyle\frac{2r-1}{2r+1}
\Bigl[\:1+\frac{8\sqrt{r}(1-r)}{(2r-1)(2r+1)}{\rm Re}(f_2^R)\:\Bigr]}
  &  {\rm for}\;\;f=b 
\end{array}
\right.
\end{equation}
with $r \equiv (M_W/m_t)^2$. Similarly we have $\alpha^{\bar{\ssf}}
=-\alpha^{\ssf}$ with replacement $f_2^R \to \bar{f}_2^L$. It should
be emphasized here that the above result means that there are no
corrections to the ``polarization vector'' $n^{\ell}_\mu$ for the
semileptonic top-quark decay. On the other hand, one can see that the
corrections to $\alpha^b$ could be substantial as the kinematical
suppression factor in the leading term $2r-1(=-0.56$) could be
canceled by the appropriate contribution from the non-standard form
factor $f_2^R$.

Applying the strategy described above and adopting the general formula
for the $t\bar{t}$ distribution $d\sigma(s_+,s_-)/d{\mit\Omega}_t$
from refs.\cite{BGH,GH_pol}, one obtains the following result for the
double distribution of the angle and the rescaled energy of $f$ for
longitudinally polarized $\epem$ beams:
\begin{equation}
\frac{d^2\sigma^{(*)}}{d\xf d\cos\thf}
=\frac{3\pi\beta\alpha_{\mbox{\tiny EM}}^2}{2s}B_{\ssf}\:
\Bigl[\:{\mit\Theta}_0^{\ssf(*)}(\xf)
+\cos\thf\,{\mit\Theta}_1^{\ssf(*)}(\xf)
+\cos^2\thf\,{\mit\Theta}_2^{\ssf(*)}(\xf) \:\Bigr],
\label{dis1}
\end{equation}
where $\alpha_{\mbox{\tiny EM}}$ is the fine structure constant and
$B_{\ssf}$ denotes the appropriate branching fraction. The energy
dependence is specified by the functions ${\mit\Theta}_i^{\ssf(*)}
(\xf)$, explicit forms of which for unpolarized beams were shown in
ref.~\cite{GH_00}.\footnote{The functions
  ${\mit\Theta}_i^{\ssf(*)}(\xf)$ for polarized beams could be
  easily obtained from formulas for unpolarized beams replacing  
  $D_{V,A,V\!\!A}$, $E_{V,A,V\!\!A}$, $F_{1\sim 4}$, $G_{1\sim 4}$
  defined by eq.(\ref{coefficients}) with $D_{V,A,V\!\!A}^{(*)}$,
  $E_{V,A,V\!\!A}^{(*)}$, $F_{1\sim 4}^{(*)}$, $G_{1\sim 4}^{(*)}$
  as in eq.(\ref{pol_coeff}) in the appendix.}\ 
They are parameterized both by the production and the decay form
factors.

The angular distribution for $f$ could be easy obtained from
eq.(\ref{dis1}) by the integration over the energy of $f$:
\begin{equation}
\frac{d\sigma^{(*)}}{d\cos\thf}
\equiv \int_{x_-}^{x_+}d\xf \frac{d^2\sigma^{(*)}}{d\xf d\cos\thf}=
\frac{3\pi\beta\alpha_{\mbox{\tiny EM}}^2}{2s}B_{\ssf}
\left({\mit\Omega}_0^{\ssf(*)}+{\mit\Omega}_1^{\ssf(*)}
\cos\thf+{\mit\Omega}_2^{\ssf(*)}\cos^2\thf\right),   
\label{dis2}
\end{equation}
where ${\mit\Omega}_i^{\ssf(*)}=\int_{x_-}^{x_+}dx
{\mit\Theta}_i^{\ssf(*)}$ are shown by eq.(\ref{omegas}) in the
appendix and $x_\pm$ define kinematical energy range of $x$:
\begin{equation}
r(1-\beta)/(1+\beta) \leq x_{\ssl} \leq 1\ \ \ \ {\rm and}
\ \ \ \ (1-r)(1-\beta)/(1+\beta) \leq x_b \leq 1-r.
\end{equation}
The decay vertex is entering the double distribution, eq.(\ref{dis1}),
through {\it i}) the functions $F^{\ssf}(\xf)$,
$G^{\ssf}(\xf)$ and $H_{1,2}^{\ssf}(\xf)$ defined in the appendix,
and {\it ii})  the depolarization factor $\alpha^{\ssf}$. All the
non-SM parts of $F^{\ssf}$, $G^{\ssf}$ and $H_{1,2}^{\ssf}$ disappear
upon integration over the energy $\xf$ both for ${\ell}^+$ and $b$, as
it could be seen from the explicit forms for
${\mit\Omega}_i^{\ssf(*)}$. Since $\alpha^{\ssf}=1$ for the leptonic
distribution, we observe that the total dependence of the lepton
distribution on non-standard structure of the top-quark decay vertex
drops out through the integration over the energy
\cite{GH_00}.\footnote{The same conclusion has also been reached
  through a different approach using the helicity formalism in
  ref.\cite{Rin00}.}\
However, one can expect substantial modifications for the
bottom-quark distribution since corrections to $\alpha^b$ could be
large.   

The fact that the angular leptonic distribution is insensitive to
corrections to the $V-A$ structure of the decay vertex allows for much
more clear tests of the production vertices through measurements of
the distribution, since that way we can avoid a contamination from a
non-standard structure of the decay vertex. As an application of the
angular distribution let us consider the following $C\!P$-violating
forward-backward charge asymmetry:\footnote{Which is an integrated
  version of the asymmetry we have considered in ref.\cite{GH_00}.}
\begin{equation}
{\cal A}_{\sst{C\!P}}^{\ssf}(P_{e^-},P_{e^+})= \frac{
{\displaystyle \int_{-c_m}^{0}\!d\cos\thf
 \frac{d\sigma^{+(*)}(\thf)}{d\cos\thf}
 -\int_{0}^{+c_m}\!d\cos\thf \frac{d\sigma^{-(*)}(\thf)}{d\cos\thf}}}
{{\displaystyle \int_{-c_m}^{0}\!d\cos\thf
  \frac{d\sigma^{+(*)}(\thf)}{d\cos\thf}
 +\int_{0}^{+c_m}\!d\cos\thf \frac{d\sigma^{-(*)}(\thf)}{d\cos\thf}}}, 
\label{asym}
\end{equation}
where $P_{e^-}$ and $P_{e^+}$ are the polarizations of $e$ and
$\bar{e}$ beams, $d\sigma^{+/-(*)}$ is referring to $f$ and $\bar{f}$
distributions respectively, and $c_m$ expresses the experimental
polar-angle cut. As $\thf \to \pi-\theta_{\bar{\ssf}}$ under $C\!P$,
this asymmetry is a true measure of $C\!P$ violation. Since
$d\sigma^{-(*)}/d\cos\thf$ is obtained from $d\sigma^{+(*)}/d\cos\thf$
by reversing the sign of $\cos\thf$ and $F_{1,4}^{(*)}$ terms and
replacing $\alpha^{\ssf}$ with $-\alpha^{\bar{\ssf}}$ in
${\mit\Omega}^{\ssf(*)}_{0,1,2}$, the asymmetry is explicitly given by
the following formula
\begin{equation}
{\cal A}_{\sst{C\!P}}^{\ssf}=N_A^{\ssf}/D_A^{\ssf}
\label{asres}
\end{equation}
with (in the leading order) 
\begin{eqnarray}
&&N_A^{\ssf}=
2c_m\alpha_0^{\ssf} \Bigl[\:(1-c_m^2){\rm Re}(F_1^{(*)})
+c_m{\rm Re}(F_4^{(*)})\:\Bigr]
\Bigl[\:1-\frac{1-\beta^2}{2\beta}\ln\frac{1+\beta}{1-\beta}\:\Bigr]
\non\\
&&\ \ \ \ \
-c_m(1-\beta^2)\alpha_1^{\ssf} {\rm Re}(f_2^R-\bar{f}_2^L) \non\\
&&\ \ \ \ \ \ \ \ \ 
\times\Bigl\{2(1-c_m^2){\rm Re}(D_{V\!\!A}^{(0,*)})+c_m E_A^{(0,*)}
\non\\
&&\ \ \ \ \ \ \ \ \ \ \
-\Bigl[\:2(1-c_m^2){\rm Re}(D_{V\!\!A}^{(0,*)})
+c_m(E_V^{(0,*)}+E_A^{(0,*)})\:\Bigr]
\frac1{2\beta}\ln\frac{1+\beta}{1-\beta} \Bigr\} \non\\
&&D_A^{\ssf}
=2c_m\Bigl[\:1+c_m^2\Bigl(1-\frac23\beta^2\Bigr)\:\Bigr]D_V^{(0,*)}
-2c_m\Bigl[\:(1-2\beta^2)-c_m^2\Bigl(1-\frac23\beta^2\Bigr)\:\Bigr]
D_A^{(0,*)} \non\\
&&\ \ \ \ \
-4c_m(1-c_m^2)\alpha_0^{\ssf}(1-\beta^2){\rm Re}(D_{V\!\!A}^{(0,*)})
\non\\
&&\ \ \ \ \
-2c_m^2[\:\alpha_0^{\ssf}(1-\beta^2)E_A^{(0,*)}
+2{\rm Re}(E_{V\!\!A}^{(0,*)})\:]
\non\\
&&\ \ \ \ \
+c_m\Bigl\{(1-c_m^2)[\:D_V^{(0,*)}+D_A^{(0,*)}
+2\alpha_0^{\ssf} {\rm Re}(D_{V\!\!A}^{(0,*)})\:] \non\\
&&\ \ \ \ \ \ \ \ \
+c_m[\:\alpha_0^{\ssf}(E_V^{(0,*)}
+E_A^{(0,*)})+2{\rm Re}(E_{V\!\!A}^{(0,*)})\:]
\Bigr\} \frac{1-\beta^2}{\beta}\ln\frac{1+\beta}{1-\beta},
\label{asexp}
\end{eqnarray}
where all the coefficients are specified in the appendix, the
superscript $(0)$ indicates the SM contribution and we expressed
$\alpha^{\ssf}$ as $\alpha^{\ssf}_0 +\alpha^{\ssf}_1 {\rm Re}(f_2^R)$
with
\begin{eqnarray*}
&&\alpha^{\ssf}_0=1,\ \ \ \ \ \ \ \ \ \ \ \
  \alpha^{\ssf}_1=0\ \ \ \ \ \ \ \ \ \ \ \ \ \ \ \ \,
  ({\rm for}\ f={\ell}),\\
&&\alpha^{\ssf}_0=\frac{2r-1}{2r+1},\ \ \ \ \;
  \alpha^{\ssf}_1=\frac{8\sqrt{r}(1-r)}{(1+2r)^2}\ \ \ ({\rm for}\ 
  f=b).
\end{eqnarray*}

As one could have anticipated, the asymmetry for $f={\ell}$ is
sensitive to $C\!P$ violation originating exclusively from the
production mechanism: It depends only on $F_{1,4}^{(*)}$ that contains
contributions from the $C\!P$-violating form factors
$\delta\!D_\gamma$ and $\delta\!D_Z$ while the contributing
decay-vertex part consists of SM $C\!P$-conserving couplings only. For
bottom quarks the effect of the modification of the decay vertex is
contained in the corrections to $b$ and $\bar{b}$ depolarization
factors, $\alpha^b+\alpha^{\bar{b}}=\alpha_1^b
{\rm Re}(f_2^R-\bar{f}_2^L)$, with SM $C\!P$-conserving contributions
from the production process.\footnote{One can show that
  $f_1^{L,R}=\pm\bar{f}_1^{L,R}$ and $f_2^{L,R}=\pm\bar{f}_2^{R,L}$
  where upper (lower) signs are those for $C\!P$-conserving
  (-violating) contributions \cite{cprelation}. Therefore, 
  when only linear terms in non-standard form factors are kept, any
  $C\!P$-violating observable defined for the top-quark decay must be
  proportional to $f_1^{L,R}-\bar{f}_1^{L,R}$ or $f_2^{L,R}-
  \bar{f}_2^{R,L}$.}

It will be instructive to give the following remark here: The
asymmetry is defined for various initial beam polarizations
$P_{e^\pm}$. For $P_{e^-} \neq  P_{e^+}$, the initial state seems not
to be $C\!P$ invariant and therefore one might expect contributions to
the asymmetry originating from the $C\!P$-conserving part of the
top-quark couplings. However, as it is seen from eq.(\ref{asexp}),
this is not the case. It turns out that even for $P_{e^-} \neq
P_{e^+}$ the asymmetry is still proportional only to the
$C\!P$-violating couplings embedded in $F_{1,2,3,4}$. The explanation
is the following: Whatever the polarizations of the initial beams are,
the electron (positron) beam consists of $e(\pm 1) (\bar{e}(\pm 1))$
where $\pm 1$ indicates the helicity, and only $e(\pm 1)$ and
$\bar{e}(\mp 1)$ can interact non-trivially in the limit of $m_e=0$
since they couple to vector bosons. Therefore the interacting initial
states are always $C\!P$ invariant.
  
Now, since we have observed in ref.\cite{GH_00} that the differential
version of the asymmetry discussed here could be substantial for
higher collider energy, in order to illustrate the potential power of
the asymmetry we present in tabs.\ref{ACPl} and \ref{ACPb} (as a
function of $\sqrt{s}$) the expected statistical significance
($N_{\!\sst{S\!D}}$) for the
asymmetry:
\renewcommand{\arraystretch}{1.4}
\begin{table}[h]
\vspace*{1.5cm}
\hspace*{1cm}(1) $P_{e^-}=P_{e^+}=0$ \vspace*{-0.25cm}
\begin{center}
\begin{tabular}{||c||c|c|c|c||} 
\hline
$\sqrt{s}\:$(GeV) & 500 & 700 & 1000 & 1500 \\ \hline
${\cal A}_{\sst{C\!P}}^{\ssl}$
& $-1.2\cdot 10^{-2}$ & $-2.6\cdot 10^{-2}$
& $-4.0\cdot 10^{-2}$ & $-5.4\cdot 10^{-2}$ \\ \hline
$N_{\!\sst{S\!D}}$ & 2.42 & 3.87 & 4.32 & 3.94      \\ \hline
\end{tabular}
\end{center}
\hspace*{1cm}(2) $P_{e^-}=P_{e^+}=+0.8$ \vspace*{-0.25cm}
\begin{center}
\begin{tabular}{||c||c|c|c|c||} 
\hline
$\sqrt{s}\:$(GeV) & 500 & 700 & 1000 & 1500 \\ \hline
${\cal A}_{\sst{C\!P}}^{\ssl}$
& $-1.4\cdot 10^{-2}$ & $-2.6\cdot 10^{-2}$
& $-3.6\cdot 10^{-2}$ & $-4.2\cdot 10^{-2}$ \\ \hline
$N_{\!\sst{S\!D}}$ & 2.80 & 4.09 & 4.04 & 3.25      \\ \hline
\end{tabular}
\end{center}
\hspace*{1cm}(3) $P_{e^-}=P_{e^+}=-0.8$ \vspace*{-0.25cm}
\begin{center}
\begin{tabular}{||c||c|c|c|c||} 
\hline
$\sqrt{s}\:$(GeV) & 500 & 700 & 1000 & 1500 \\ \hline
${\cal A}_{\sst{C\!P}}^{\ssl}$
& $-1.2\cdot 10^{-2}$ & $-2.6\cdot 10^{-2}$
& $-4.1\cdot 10^{-2}$ & $-5.7\cdot 10^{-2}$ \\ \hline
$N_{\!\sst{S\!D}}$ & 3.53 & 5.72 & 6.57 & 6.20      \\ \hline
\end{tabular}
\end{center}
\caption{The $C\!P$-violating asymmetry ${\cal A}_{\sst{C\!P}}^{\ssl}$
and the expected statistical significance $N_{\!\sst{S\!D}}$ for ${\rm Re}
(\delta\!D_{\gamma,Z})=+0.05$, and beam polarizations
$P_{e^-}=P_{e^+}=(1)\ 0$, (2) $+0.8$ and (3) $-0.8$ as an example.}
\label{ACPl}
\end{table}
\newpage
\begin{table}[h]
\hspace*{1cm}(1) $P_{e^-}=P_{e^+}=0$ \vspace*{-0.25cm}
\begin{center}
\begin{tabular}{||c||c|c|c|c||} 
\hline
$\sqrt{s}\:$(GeV) & 500 & 700 & 1000 & 1500 \\ \hline
${\cal A}_{\sst{C\!P}}^{\ssb}$
& $+1.2\cdot 10^{-2}$ & $+1.7\cdot 10^{-2}$
& $+2.2\cdot 10^{-2}$ & $+2.6\cdot 10^{-2}$ \\ \hline
$N_{\!\sst{S\!D}}$ & 5.10 & 5.50 & 5.03 & 4.03      \\ \hline
\end{tabular}
\end{center}
\hspace*{1cm}(2) $P_{e^-}=P_{e^+}=+0.8$ \vspace*{-0.25cm}
\begin{center}
\begin{tabular}{||c||c|c|c|c||} 
\hline
$\sqrt{s}\:$(GeV) & 500 & 700 & 1000 & 1500 \\ \hline
${\cal A}_{\sst{C\!P}}^{\ssb}$
& $-9.4\cdot 10^{-3}$ & $-4.6\cdot 10^{-3}$
& $+1.4\cdot 10^{-3}$ & $+7.8\cdot 10^{-3}$ \\ \hline
$N_{\!\sst{S\!D}}$ & 4.04 & 1.52 & 0.33 & 1.27      \\ \hline
\end{tabular}
\end{center}
\hspace*{1cm}(3) $P_{e^-}=P_{e^+}=-0.8$ \vspace*{-0.25cm}
\begin{center}
\begin{tabular}{||c||c|c|c|c||} 
\hline
$\sqrt{s}\:$(GeV) & 500 & 700 & 1000 & 1500 \\ \hline
${\cal A}_{\sst{C\!P}}^{\ssb}$
& $+2.6\cdot 10^{-2}$ & $+3.0\cdot 10^{-2}$
& $+3.3\cdot 10^{-2}$ & $+3.5\cdot 10^{-2}$ \\ \hline
$N_{\!\sst{S\!D}}$ & 16.0 & 14.3 & 11.2 & 8.02      \\ \hline
\end{tabular}
\end{center}
\caption{The $C\!P$-violating asymmetry ${\cal A}_{\sst{C\!P}}^{\ssb}$
and the expected statistical significance $N_{\!\sst{S\!D}}$ for ${\rm Re}
(\delta\!D_{\gamma,Z})={\rm Re}(f_2^R-\bar{f}_2^L)=+0.05$, 
and beam
polarizations $P_{e^-}=P_{e^+}=(1)\ 0$, (2) $+0.8$ and (3) $-0.8$
as an example.}
\label{ACPb}
\end{table}
\begin{equation}
N_{\!\sst{S\!D}}\equiv
\frac{|{\cal A}_{\sst{C\!P}}^{\ssf}|}
{{\mit\Delta}{\cal A}_{\sst{C\!P}}^{\ssf}}
=|{\cal A}_{\sst{C\!P}}^{\ssf}|
\sqrt{\smash{\frac{L_{e\!f\!f}\sigma_{tot}}
{1-({\cal A}_{\sst{C\!P}}^{\ssf})^2}}\vphantom{A^2\over A}},
\end{equation}
where $L_{e\!f\!f}\equiv \epsilon L$ is an effective integrated
luminosity for the tagging efficiency $\epsilon$. Hereafter we adopt
the integrated luminosity $L=500\fbarn^{-1}$ and the efficiency
$\epsilon=60\%$ both for lepton and $b$-quark detection.\footnote{That
  low efficiency is supposed to take into account cuts necessary to
  suppress the background. If the $b$-tagging is applied then, as
  shown in the second paper of ref.\cite{background}, the irreducible
  background to top events due $W^\pm \; + 2\;b\;+2\;j$ is negligible,
  provided that a vertex tagging efficiency $\epsilon_b \gesim .5$ can
  be achieved. Therefore for the $b$-tagging case the efficiency we
  have employed is definitely conservative. Since $N_{\!\sst{S\!D}}$
  scales as $\sqrt{\epsilon L}$ it would be easy to estimate the
  statistical significance for any given luminosity and efficiency.}\ 
In addition, to fit the typical
detector shape \cite{fujii} we impose a polar-angle cut $|\cos\thf|<
0.9$, i.e. $c_m=0.9$ in eq.(\ref{asexp}), both for leptons and bottom
quarks. On the other hand, we will not impose any cut on the lepton/
$b$-quark energy since their kinematical lower bounds $E_{\ssl}^{min}=
7.5$ GeV and $E_{\ssb}^{min}=27.5$ GeV (for $\sqrt{s}=500$ GeV) are
large enough to be detected. Perfect angular resolution will be
assumed both for lepton and $b$-quark final states. Also ideal
leptonic-energy resolution will be used.

As it is seen from the tables, the asymmetry
${\cal A}_{\sst{C\!P}}^{\ssf}$ turned out to be a very sensitive
$C\!P$-violating observable; even for unpolarized beams and
$C\!P$-violating couplings of the order of $0.05$ one can expect
$2.4\sigma\sim 5.5 \sigma$ effect both for lepton and $b$-quark
asymmetries once $L=500\fbarn^{-1}$ is achieved.

The $C\!P$-violating form factors discussed here could be also
generated within the SM. However, it is easy to notice that the first
non-zero contribution to $\delta\!D_{\gamma,Z}$ would require at least
two loops. For the top-quark decay process $C\!P$ violation could
appear at the one-loop level, however it is strongly suppressed by the
double GIM mechanism \cite{GK}. Therefore we can conclude that an
experimental detection of $C\!P$-violating form factors considered
here would be a clear indication for physics beyond the SM. In
particular, non-vanishing ${\cal A}_{\sst{C\!P}}^l$ in the lepton
distribution will strongly indicate some new-physics in
$t\bar{t}\gamma/Z$ couplings.

\sec{Optimal-Observable Analysis}
\setcounter{equation}{0}

\noindent 
{\bf 4.1. Optimal observables}

\noindent
Let us briefly recall the main points of the optimal-observable (OO)
technique \cite{opt}. Suppose we have a distribution
\begin{equation}
\frac{d\sigma}{d\phi}(\equiv{\mit\Sigma}(\phi))=\sum_i c_i f_i(\phi)
\label{dsdphi}
\end{equation}
where $f_i(\phi)$ are known functions of the location in final-state
phase space $\phi$ and $c_i$'s are model-dependent coefficients. The
goal would be to determine  $c_i$'s. It can be done by using
appropriate weighting functions $w_i(\phi)$ such that $\int d\phi
w_i(\phi){\mit\Sigma}(\phi)=c_i$. Generally, different choices for
$w_i(\phi)$ are possible, but there is a unique choice so that the
resultant statistical error is minimized. Such functions are given by
\begin{equation}
w_i(\phi)=\sum_j X_{ij}f_j(\phi)/{\mit\Sigma}(\phi)\,, \label{X_def}
\end{equation}
where $X_{ij}$ is the inverse matrix of $M_{ij}$ which is defined as
\begin{equation}
M_{ij}\equiv \int d\phi{f_i(\phi)f_j(\phi)\over{\mit\Sigma}(\phi)}\,.
\label{M_def}
\end{equation}
The statistical uncertainty of $c_i$-\-deter\-mination through
$d\sigma/d\phi$ measurement becomes
\begin{equation}
{\mit\Delta}c_i=\sqrt{X_{ii}\,\sigma_T/N}\,, \label{delc_i}
\end{equation}
where $\sigma_T\equiv\int d\phi(d\sigma/d\phi)$ and $N$ is the total
number of events.

It is clear from the definition of the matrix $M_{ij}$,
eq.(\ref{M_def}), that $M_{ij}$ has no inverse if the functions
$f_i(\phi)$ are linearly dependent, and then we cannot perform any
meaningful analysis. One can see it more intuitively as follows: if
$f_i(\phi)=f_j(\phi)$ the splitting between $c_i$ and $c_j$ would be
totally arbitrary and only $c_i+c_j$ could be determined.

\noindent
{\bf 4.2. For application} 

\noindent
In order to apply the OO procedure to the processes under
consideration, we have to reexpress the distributions in the form
shown in eq.(\ref{dsdphi}). The angular distribution, eq.(\ref{dis2}),
has already an appropriate form for this purpose, where $f_i(\phi)=
\cos^i\thf$ $(i=0,1,2)$ and ${\mit\Omega}^{\ssf(*)}_i$ are the
coefficients to be determined. On the other hand, the double angular
and energy distribution eq.(\ref{dis1}) must be modified. We reexpress
the distribution in the following way, keeping only the SM
contribution and terms linear in the non-standard form factors:
\begin{equation}
\frac{d^2\sigma^{(*)}}{d\xf d\cos\thf}
=\frac{3\pi\beta\alpha_{\mbox{\tiny EM}}^2}{2s}B_{\!\ssf}\:S^{(*)}_{\!\ssf}(\xf, \thf),
\label{reexp}
\end{equation}
where 
\begin{eqnarray*}
&&S^{(*)}_{\!\ssf}(\xf, \thf)=S^{(0,*)}_{\!\ssf}(\xf, \thf) \\
&&\phantom{S^{(*)}(\xf, \thf)=}
+\sum_{v=\gamma,Z}\Bigl[\:
 {\rm Re}(\delta\!A_v){\cal F}_{Av}^{\ssf(*)}(\xf, \thf)
+{\rm Re}(\delta\!B_v){\cal F}_{Bv}^{\ssf(*)}(\xf, \thf) \\
&&\phantom{S^{(*)}(\xf, \thf)=\sum}
+{\rm Re}(\delta  C_v){\cal F}_{Cv}^{\ssf(*)}(\xf, \thf)
+{\rm Re}(\delta\!D_v){\cal F}_{Dv}^{\ssf(*)}(\xf, \thf)
\:\Bigr] \\
&&\phantom{S^{(*)}(\xf, \thf)=}
+{\rm Re}(f^R_2){\cal F}_{2R}^{\ssf(*)}(\xf, \thf).
\end{eqnarray*}

As it is seen from the above formula, the coefficients $c_i$ of
eq.(\ref{dsdphi}) are just the anomalous form factors to be determined.
The SM contribution reads:
\begin{equation}
S^{(0,*)}_{\!\ssf}(\xf, \thf) ={\mit\Theta}^{\ssf(0,*)}_0(\xf)
+\cos\thf\,{\mit\Theta}^{\ssf(0,*)}_1(\xf)
+\cos^2\thf\,{\mit\Theta}^{\ssf(0,*)}_2(\xf)
\end{equation}
with
\begin{eqnarray}
{\mit\Theta}^{\ssf(0,*)}_0(x)\!\!\!&
=\!\!\!&\frac12\Bigl[\:(3-\beta^2)D_V^{(0,*)}-(1-3\beta^2)D_A^{(0,*)} 
-2\alpha_0^{\ssf}(1-\beta^2){\rm Re}(D_{V\!\!A}^{(0,*)}) 
     \:\Bigr]\:f^{\ssf}(x) \non\\
&&+2\alpha_0^{\ssf}{\rm Re}(D_{V\!\!A}^{(0,*)})\:g^{\ssf}(x) \non\\
&&+\frac12\Bigl[\:D_V^{(0,*)}+D_A^{(0,*)}
     +2\alpha_0^{\ssf}\:{\rm Re}(D_{V\!\!A}^{(0,*)})\:\Bigr]\:
   \Bigl[\:2h_1^{\ssf}(x)-h_2^{\ssf}(x)\:\Bigr], \\
{\mit\Theta}^{\ssf(0,*)}_1(x)\!\!\!&=
\!\!\!&2\Bigl[\:2{\rm Re}(E_{V\!\!A}^{(0,*)})
+\alpha_0^{\ssf}(1-\beta^2)E_A^{(0,*)}\:\Bigr]f^{\ssf}(x) 
+2\alpha_0^{\ssf}\:(E_V^{(0,*)}+E_A^{(0,*)})\:
     g^{\ssf}(x) \non\\
&&-2\:\Bigl[\:2{\rm Re}(E_{V\!\!A}^{(0,*)})
   +\alpha_0^{\ssf}(\:E_V^{(0,*)}+
E_A^{(0,*)}\:)\:\Bigr]\:h_1^{\ssf}(x), \\
{\mit\Theta}^{\ssf(0,*)}_2(x)\!\!\!&=\!\!\!&
   \frac12\Bigl[\:(3-\beta^2)(D_V^{(0,*)}+D_A^{(0,*)})
   +6\alpha_0^{\ssf}(1-\beta^2){\rm Re}(D_{V\!\!A}^{(0,*)})
   \:\Bigr]\:f^{\ssf}(x) \non\\
&&+2\alpha_0^{\ssf}{\rm Re}(D_{V\!\!A}^{(0,*)})\:g^{\ssf}(x) \non\\
&&-\frac32\Bigl[\:D_V^{(0,*)}+D_A^{(0,*)}
    +2\alpha_0^{\ssf}{\rm Re}(D_{V\!\!A}^{(0,*)})\:\Bigr]\:
  \Bigl[\:2h_1^{\ssf}(x)-h_2^{\ssf}(x) \:\Bigr].
\end{eqnarray}
\noindent
Explicit forms of the functions 
${\cal F}^{f(*)}_{\{A,B,C,D\}\{\gamma,Z\}}$ and
${\cal F}^{f(*)}_{2R}$ are shown in the appendix together with the
functions $f^{\ssf}(x)$, $g^{\ssf}(x)$ and $h_{1,2}^{\ssf}(x)$.

There are ten functions entering eq.(\ref{reexp}): $S^{(0,*)}_{\ssf}$,
${\cal F}^{f(*)}_{\{A,B,C,D\}\{\gamma,Z\}}$ and
${\cal F}^{f(*)}_{2R}$. As explained earlier, one cannot determine
their coefficients separately if they are not independent. As could be
found from the appendix, for the double lepton distribution, the first
nine functions are linear combinations of
\begin{eqnarray}
&&f^{\ssl}(x),\ \ f^{\ssl}(x)\cos\theta,\ \ f^{\ssl}(x)\cos^2\theta,
\non \\
&&g^{\ssl}(x),\ \ g^{\ssl}(x)\cos\theta,\ \ g^{\ssl}(x)\cos^2\theta,
\non \\
&&h^{\ssl}_{1,2}(x)(1-3\cos^2\theta),\ \ \ h^{\ssl}_1(x)\cos\theta,
\label{fghh}
\end{eqnarray}
while the last one, ${\cal F}^{\ssl(*)}_{2R}$, is a combination of
$\delta\{f^{\ssl},g^{\ssl},h^{\ssl}_{1,2}\}(x)$ and $\cos^n\theta$
($n=0,1,2$). Since there are ten coefficients to be
measured,\footnote{Counting the SM coefficient in front of
  $S^{(0,*)}_{\ssf}$ which is normalized to $1$.}\ 
it looks always possible to determine all of them. However, it turns
out not to be the case in some special cases. Indeed the possibility
for the determination of all the ten form factors depends crucially on
the chosen beam polarization.

This can be understood considering the invariant amplitude for
$e\bar{e}\to t\bar{t}$, which could be expressed in terms of eight
independent parameters as
\begin{eqnarray*}
&&{\cal M}(e\bar{e} \to t\bar{t})
=C_{VV}\:[\,\bar{v}_e \gamma_\mu u_e
       \cdot\bar{u}_t \gamma^\mu v_{\bar{t}} \,]
  +C_{V\!\!A}\:[\,\bar{v}_e \gamma_\mu u_e
       \cdot\bar{u}_t \gamma_5 \gamma^\mu v_t \,]             \\
&&\phantom{{\cal M}(e\bar{e} \to t\bar{t})}
+C_{AV}\:[\,\bar{v}_e \gamma_5 \gamma_\mu u_e
       \cdot\bar{u}_t \gamma^\mu v_t \,]
  +C_{AA}\:[\,\bar{v}_e \gamma_5 \gamma_\mu u_e
       \cdot\bar{u}_t \gamma_5 \gamma^\mu v_t \,]             \\
&&\phantom{{\cal M}(e\bar{e} \to t\bar{t})}
+C_{V\!S}\:[\,\bar{v}_e \slaq u_e
       \cdot\bar{u}_t v_t \,]
  +C_{V\!P}\:[\,\bar{v}_e \slaq u_e
       \cdot\bar{u}_t \gamma_5 v_t \,]                        \\
&&\phantom{{\cal M}(e\bar{e} \to t\bar{t})}
+C_{AS}\:[\,\bar{v}_e \gamma_5 \slaq u_e
       \cdot\bar{u}_t v_t \,]
  +C_{AP}\:[\,\bar{v}_e \gamma_5 \slaq u_e
       \cdot\bar{u}_t \gamma_5 v_t \,].
\end{eqnarray*}
However, if $e$ or $\bar{e}$ is perfectly polarized, contributions
from $[\bar{v}_e \gamma_\mu u_e]$ and $[\bar{v}_e \gamma_5\gamma_\mu
u_e]$ are identical. For example, when $e$ has fully left-handed
polarization, $u_e$ is replaced with $u_{eL} \equiv (1-\gamma_5)u_e/2$
and in that case they are changed as
\[
\bar{v}_e \gamma_\mu u_e\ \ \to\ \ \bar{v}_e \gamma_\mu u_{eL},\ \ \ \
\bar{v}_e \gamma_5\gamma_\mu u_e\ \ 
                            \to\ \ \bar{v}_e \gamma_\mu u_{eL}.
\]
Therefore, the invariant amplitude becomes
\begin{eqnarray*}
&&{\cal M}(e\bar{e} \to t\bar{t})                             \\
&&=(C_{VV}+C_{AV})\:[\,\bar{v}_e \gamma_\mu u_{eL}
       \cdot\bar{u}_t \gamma^\mu v_{\bar{t}} \,]
  +(C_{V\!\!A}+C_{AA})\:[\,\bar{v}_e \gamma_\mu u_{eL}
       \cdot\bar{u}_t \gamma_5 \gamma^\mu v_t \,]             \\
&&+(C_{V\!S}+C_{AS})\:[\,\bar{v}_e \slaq u_{eL}
       \cdot\bar{u}_t v_t \,]
  +(C_{V\!P}+C_{AP})\:[\,\bar{v}_e \slaq u_{eL}
       \cdot\bar{u}_t \gamma_5 v_t \,],
\end{eqnarray*}
and one ends with just four independent functions and therefore only
four coefficients could be determined. More details could be found in
the appendix below eq.(\ref{other-coeff}). Of course, such singular
configurations of polarization are not considered in our analyses.

\begin{figure}
\postscript{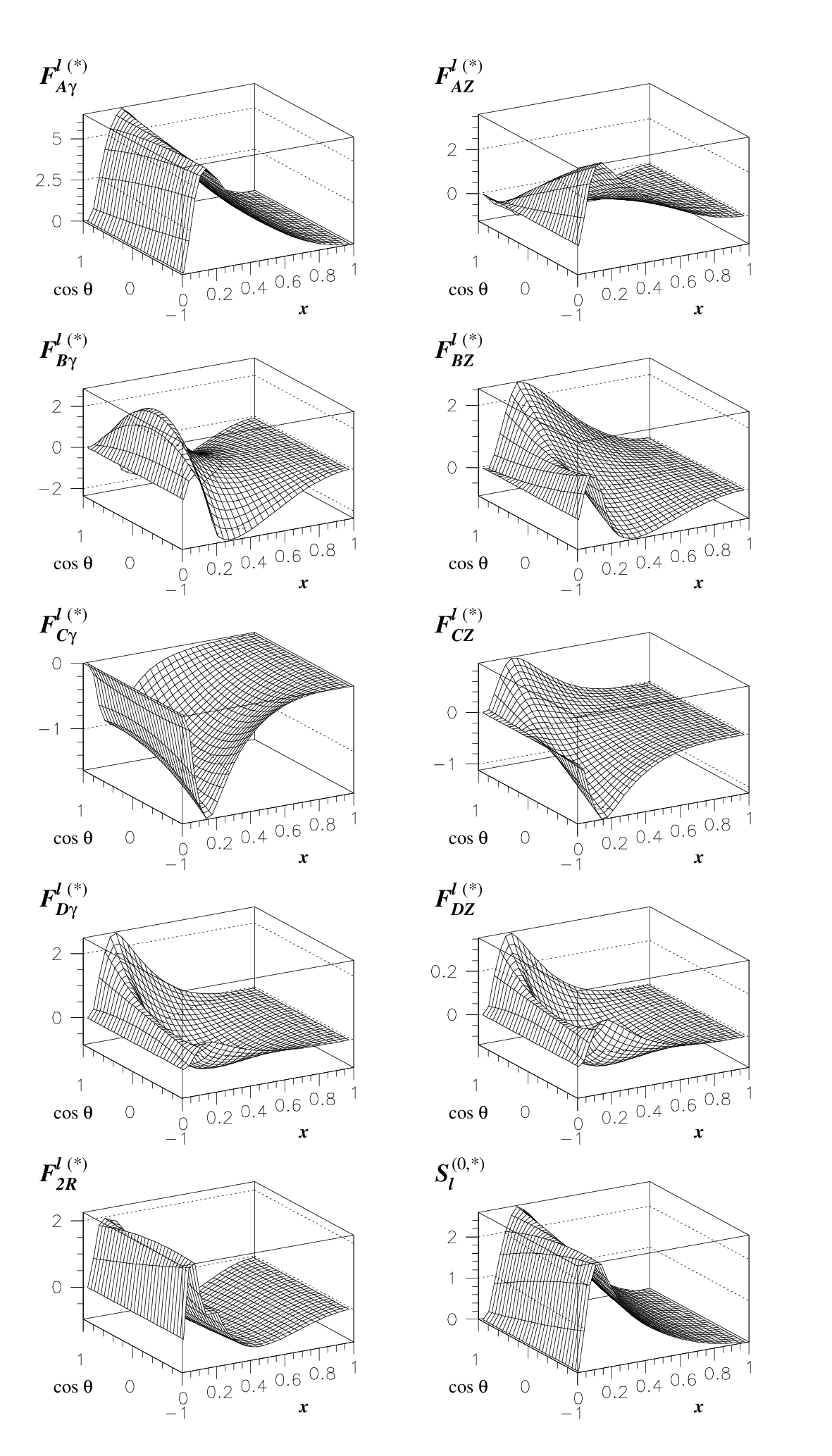}{0.8}
\caption{The shape of the coefficient functions
${\cal F}^{\ssl(*)}_{\{A,B,C,D\}\{\gamma,Z\}}$,
${\cal F}^{\ssl(*)}_{2R}$, and $S^{(0,*)}_{\ssl}$ for unpolarized
beams}
\label{co-func00}                                                              
\end{figure}
\begin{figure}
\postscript{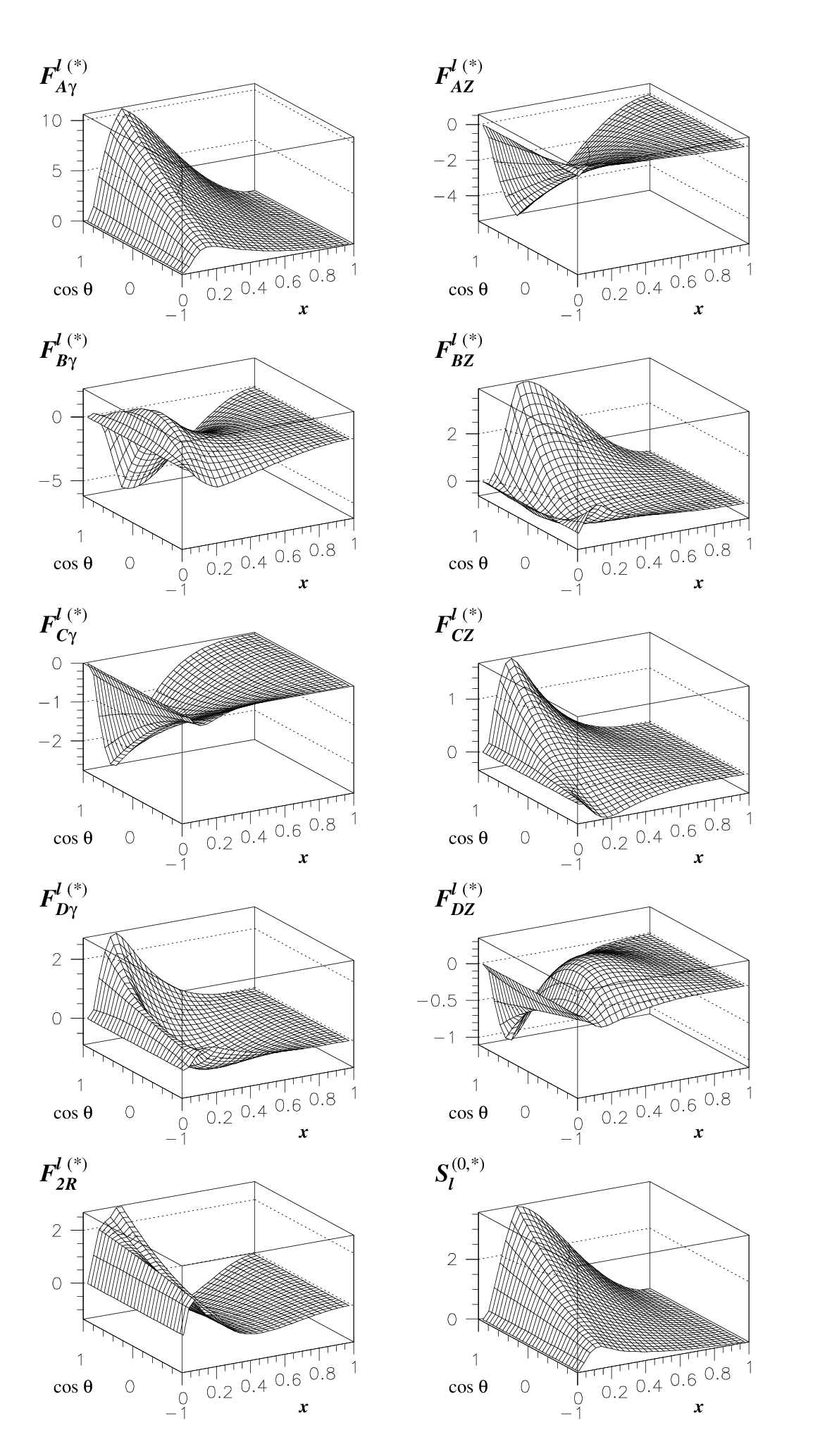}{0.8}
\caption{The shape of the coefficient functions
${\cal F}^{\ssl(*)}_{\{A,B,C,D\}\{\gamma,Z\}}$,
${\cal F}^{\ssl(*)}_{2R}$, and $S^{(0,*)}_{\ssl}$ for
$P_{e^-}=P_{e^+}=0.5$}
\label{co-func55}                                                              
\end{figure}

As for $b$-quark distributions $\delta\!f^{\ssb}(x)=\delta g^{\ssb}(x)
=\delta h_1^{\ssb}(x)=\delta h_2^{\ssb}(x)=0$, instead of ten
functions $\phi_i(x)$ we have in that case only nine of them given by
the $b$-quark version eq.(\ref{fghh}). Therefore, at most nine
couplings could be determined. Since $b$-quark energy resolution is
expected to be relatively poor, we will not apply OO procedure to the
$b$-quark double distribution.

\noindent 
{\bf 4.3. Numerical analysis}

\noindent
Below, we will adjust beam polarizations to perform the best
measurement of the form factors. In order to gain some intuition we
show in figs. \ref{co-func00} and \ref{co-func55} the functions 
${\cal F}_{\{A,B,C,D\}\{\gamma,Z\}}^{\ssl(*)}$,
${\cal F}_{2R}^{\ssl(*)}$ plus $S^{(0,*)}_{\ssl}$ for unpolarized
beams (fig.\ref{co-func00}) and for the beam polarization $P_{e^-}=
P_{e^+}=+0.5$ (fig.\ref{co-func55}). The figures illustrates how much
the polarization could modify the functions and therefore influence
the possibility for the determination of the form factors. 

\noindent 
{\bf Lepton angular distribution}

\noindent
Since we have only three independent functions $\{1,\:\cos\theta,\:
\cos^2\theta\}$, $M$ and its inverse $X$ are $(3,3)$ matrices. We have
considered the following polarization set-ups: $P_{e^-}=P_{e^+}=0,\:
\pm 0.5$  and $\pm 1$. Since $1>|\cos\theta|>\cos^2\theta$ we observe
that $X_{11}<X_{22}<X_{33}$, therefore the statistical uncertainty for 
${\mit\Omega}^{(*)}_0$ measurement, ${\mit\Delta\Omega}^{(*)}_0$, is
always the smallest one.

Once we assume the detection efficiency $\epsilon$ and the integrated
luminosity $L$, we can compute the statistical significance of
measuring the non-SM part of ${\mit\Omega}^{(*)}_i$
$$
N^{(i)}_{\!\sst{S\!D}}=|{\mit\Omega}^{(*)}_i-{\mit\Omega}^{(0,*)}_i|/
{\mit\Delta}{\mit\Omega}^{(*)}_i.
$$
For the efficiency and luminosity specified earlier we obtain\\
\noindent
$\bullet\:P_{e^-}=P_{e^+}=0$
\begin{eqnarray}
&&M_{11}=2.06,\ \ \  M_{22}=0.55,\ \ \ M_{33}=0.27 \non\\
&&X_{11}=1.09,\ \ \  X_{22}=1.83,\ \ \ X_{33}=8.40 \non\\
&&N_{\!\sst{S\!D}}^{(0)}=16.1,\ \ \ 
  N_{\!\sst{S\!D}}^{(1)}=3.5,\ \ \ 
  N_{\!\sst{S\!D}}^{(2)}=1.9\non
\end{eqnarray}
\noindent
$\bullet\:P_{e^-}=P_{e^+}=+0.5$
\begin{eqnarray}
&&M_{11}=3.06,\ \ \  M_{22}=0.95,\ \ \ M_{33}=0.49 \non\\
&&X_{11}=0.82,\ \ \  X_{22}=1.66,\ \ \ X_{33}=6.25 \non\\
&&N_{\!\sst{S\!D}}^{(0)}=7.5,\ \ \ 
  N_{\!\sst{S\!D}}^{(1)}=2.7,\ \ \ 
  N_{\!\sst{S\!D}}^{(2)}=1.3\non
\end{eqnarray}
\noindent
$\bullet\:P_{e^-}=P_{e^+}=+1$
\begin{eqnarray}
&&M_{11}=3.20,\ \ \  M_{22}=1.25,\ \ \ M_{33}=0.72 \non\\
&&X_{11}=1.00,\ \ \  X_{22}=2.39,\ \ \ X_{33}=7.00 \non\\
&&N_{\!\sst{S\!D}}^{(0)}=5.2,\ \ \ 
  N_{\!\sst{S\!D}}^{(1)}=3.0,\ \ \
  N_{\!\sst{S\!D}}^{(2)}=1.3\non
\end{eqnarray}
\noindent
$\bullet\:P_{e^-}=P_{e^+}=-0.5$
\begin{eqnarray}
&&M_{11}=1.27,\ \ \  M_{22}=0.35,\ \ \ M_{33}=0.17 \non\\
&&X_{11}=1.81,\ \ \  X_{22}=2.91,\ \ \ X_{33}=13.4 \non\\
&&N_{\!\sst{S\!D}}^{(0)}=26.2,\ \ \ 
  N_{\!\sst{S\!D}}^{(1)}=4.9,\ \ \ 
  N_{\!\sst{S\!D}}^{(2)}=3.0 \non
\end{eqnarray}
\noindent
$\bullet\:P_{e^-}=P_{e^+}=-1$
\begin{eqnarray}
&&M_{11}=0.76,\ \ \  M_{22}=0.21,\ \ \ M_{33}=0.11 \non\\
&&X_{11}=3.06,\ \ \  X_{22}=4.90,\ \ \ X_{33}=22.3 \non\\
&&N_{\!\sst{S\!D}}^{(0)}=35.5,\ \ \ 
  N_{\!\sst{S\!D}}^{(1)}=6.4,\ \ \ 
  N_{\!\sst{S\!D}}^{(2)}=4.0,
\end{eqnarray}
where we put all the non-SM parameters
${\rm Re}(\delta\{A,B,C,D\}_{\gamma,Z})$ and ${\rm Re}(f_2^R)$ to be
$+0.05$ as an example.

As one can see, the precision is better for negative beam
polarization, partly because of larger number of events. However we
cannot conclude that using negatively-polarized beams is always more
effective for new-physics search, since $N_{\!\sst{S\!D}}^{(i)}$
strongly depends on the non-SM parameters used in the computations. In
fact, positively-polarized beams give smaller $X_{ii}$ and this is
independent of the choice of non-SM parameters. Therefore polarization
of the initial beams should be carefully adjusted for each tested
model in actual experimental analysis.

\noindent 
{\bf $\mib{b}$-quark angular distribution}

\noindent
We can compute $M$, $X$ and $N_{\!\sst{S\!D}}^{(i)}$ in the same way
as for the lepton distribution:

\noindent
$\bullet\:P_{e^-}=P_{e^+}=0$
\begin{eqnarray}
&&M_{11}=2.23,\ \ \  M_{22}=0.63,\ \ \ M_{33}=0.31 \non\\
&&X_{11}=1.04,\ \ \  X_{22}=1.85,\ \ \ X_{33}=7.98 \non\\
&&N_{\!\sst{S\!D}}^{(0)}=37.3,\ \ \ 
  N_{\!\sst{S\!D}}^{(1)}=17.8,\ \ \ 
  N_{\!\sst{S\!D}}^{(2)}=1.9\non
\end{eqnarray}
\noindent
$\bullet\:P_{e^-}=P_{e^+}=+0.5$
\begin{eqnarray}
&&M_{11}=2.38,\ \ \  M_{22}=0.65,\ \ \ M_{33}=0.32 \non\\
&&X_{11}=0.96,\ \ \  X_{22}=1.61,\ \ \ X_{33}=7.29 \non\\
&&N_{\!\sst{S\!D}}^{(0)}=15.5,\ \ \ 
  N_{\!\sst{S\!D}}^{(1)}=7.6,\ \ \ 
  N_{\!\sst{S\!D}}^{(2)}=1.8\non
\end{eqnarray}
\noindent
$\bullet\:P_{e^-}=P_{e^+}=+1$
\begin{eqnarray}
&&M_{11}=1.63,\ \ \  M_{22}=0.45,\ \ \ M_{33}=0.22 \non\\
&&X_{11}=1.39,\ \ \  X_{22}=2.29,\ \ \ X_{33}=10.5 \non\\
&&N_{\!\sst{S\!D}}^{(0)}=9.7,\ \ \ 
  N_{\!\sst{S\!D}}^{(1)}=4.8,\ \ \ 
  N_{\!\sst{S\!D}}^{(2)}=2.2\non
\end{eqnarray}
\noindent
$\bullet\:P_{e^-}=P_{e^+}=-0.5$
\begin{eqnarray}
&&M_{11}=1.45,\ \ \  M_{22}=0.42,\ \ \ M_{33}=0.21 \non\\
&&X_{11}=1.63,\ \ \  X_{22}=3.01,\ \ \ X_{33}=12.5 \non\\
&&N_{\!\sst{S\!D}}^{(0)}=62.6,\ \ \ 
  N_{\!\sst{S\!D}}^{(1)}=29.3,\ \ \ 
  N_{\!\sst{S\!D}}^{(2)}=2.4\non
\end{eqnarray}
\noindent
$\bullet\:P_{e^-}=P_{e^+}=-1$
\begin{eqnarray}
&&M_{11}=0.87,\ \ \  M_{22}=0.25,\ \ \ M_{33}=0.13 \non\\
&&X_{11}=2.74,\ \ \  X_{22}=5.10,\ \ \ X_{33}=21.0 \non\\
&&N_{\!\sst{S\!D}}^{(0)}=85.1,\ \ \ 
  N_{\!\sst{S\!D}}^{(1)}=39.7,\ \ \ 
  N_{\!\sst{S\!D}}^{(2)}=3.1,
\end{eqnarray}
for ${\rm Re}(\delta\{A,B,C,D\}_{\gamma,Z})={\rm Re}(f_2^R)=+0.05$.
Negatively-polarized beams give better precision again, but the same
remark as to the lepton angular distribution should be kept in mind
also here.

The above results prove that the optimal observables utilizing the
angular distributions should be very efficient seeking for the non-SM
parts of ${\mit\Omega}_i^{\ssf(*)}$. However, since they are
combinations of the form factors, we can only constrain them. Of
course, it would be exciting if we found any signal of non-standard
physics, however our final goal is to determine each form factor
separately. That is why we proceed to the next analysis using the
double angular and energy distributions.

\noindent 
{\bf Lepton angular and energy distribution}

\noindent
Because of high precision of direction and energy determination of
leptons we adopted the double energy and angular distributions,
eq.(\ref{reexp}), also for OO analysis. As discussed earlier, in
principle all nine form factors could be determined with the expected
statistical uncertainties ${\mit\Delta}c_i$ for $c_i={\rm Re}
(\delta\{A,B,C,D\}_{\gamma,Z})$ and ${\rm Re}(f_2^R)$. The beam
polarizations $P_{e^-}$ and $P_{e^+}$ were adjusted to minimize the
statistical error for determination of each form factor. We found that
positive polarizations lead to a smaller ${\mit\Delta}c_i$ for eight
form factors in the production vertices. Unfortunately, however, the
optimal polarizations for each form-factor measurement is different.
Below we present the smallest statistical uncertainties and the
corresponding beam polarizations for each parameter:
\begin{eqnarray}
&&{\mit\Delta}[\:\re(\delta\!A_\gamma)\:]=0.16\ \ \ \ \ \ \ 
  {\rm for}\ P_{e^-}=0.7\ {\rm and}\ P_{e^+}=0.7, \non\\
&&{\ \ \ \
\Bigl(^{\ \delta\!A_Z:     \:0.13,\ \delta\!B_\gamma:\:0.25,\
          \delta\!B_Z:     \:0.49,\ \delta  C_\gamma:\:2.47 \ }_{\ 
          \delta  C_Z:     \:4.69,\ \delta\!D_\gamma:\:27.2,\
          \delta\!D_Z:     \:53.2,\:\phantom{\delta}f_2^R:\:0.02 \
                   }\Bigr),} \non\\
&&{\mit\Delta}[\:\re(\delta\!A_Z)\:]=0.07\ \ \ \ \ \ \ 
  {\rm for}\ P_{e^-}=0.5\ {\rm and}\ P_{e^+}=0.4, \non\\
&&{\ \ \ \
\Bigl(^{\ \delta\!A_\gamma:\:0.23,\ \delta\!B_\gamma:\:0.11,\
          \delta\!B_Z:     \:0.27,\ \delta  C_\gamma:\:0.70 \ }_{\ 
          \delta  C_Z:     \:1.76,\ \delta\!D_\gamma:\:7.09,\
          \delta\!D_Z:     \:20.6,\:\phantom{\delta}f_2^R:\:0.02 \
                   }\Bigr),} \non\\
&&{\mit\Delta}[\:\re(\delta\!B_\gamma)\:]=0.09\ \ \ \ \ \ \ 
  {\rm for}\ P_{e^-}=0.2\ {\rm and}\ P_{e^+}=0.2, \non\\
&&{\ \ \ \
\Bigl(^{\ \delta\!A_\gamma:\:0.43,\ \delta\!A_Z:     \:0.11,\
          \delta\!B_Z:     \:0.36,\ \delta  C_\gamma:\:0.21 \ }_{\ 
          \delta  C_Z:     \:1.17,\ \delta\!D_\gamma:\:0.95,\
          \delta\!D_Z:     \:14.6,\:\phantom{\delta}f_2^R:\:0.03 \
                   }\Bigr),} \non\\
&&{\mit\Delta}[\:\re(\delta\!B_Z)\:]=0.27\ \ \ \ \ \ \ 
  {\rm for}\ P_{e^-}=0.4\ {\rm and}\ P_{e^+}=0.4, \non\\
&&{\ \ \ \
\Bigl(^{\ \delta\!A_\gamma:\:0.25,\ \delta\!A_Z:     \:0.07,\
          \delta\!B_\gamma:\:0.10,\ \delta  C_\gamma:\:0.56 \ }_{\ 
          \delta  C_Z:     \:1.56,\ \delta\!D_\gamma:\:5.43,\
          \delta\!D_Z:     \:18.5,\:\phantom{\delta}f_2^R:\:0.02 \
                   }\Bigr),} \non\\
&&{\mit\Delta}[\:\re(\delta C_\gamma)\:]=0.11\ \ \ \ \ \ \ 
  {\rm for}\ P_{e^-}=0.1\ {\rm and}\ P_{e^+}=0.0, \non\\
&&{\ \ \ \
\Bigl(^{\ \delta\!A_\gamma:\:0.82,\ \delta\!A_Z:     \:0.22,\
          \delta\!B_\gamma:\:0.10,\ \delta\!B_Z:     \:0.65 \ }_{\ 
          \delta  C_Z:     \:1.11,\ \delta\!D_\gamma:\:1.76,\
          \delta\!D_Z:     \:14.6,\:\phantom{\delta}f_2^R:\:0.03 \
                   }\Bigr),} \non\\
&&{\mit\Delta}[\:\re(\delta C_Z)\:]=1.11\ \ \ \ \ \ \ 
  {\rm for}\ P_{e^-}=0.1\ {\rm and}\ P_{e^+}=0.0, \non\\
&&{\ \ \ \
\Bigl(^{\ \delta\!A_\gamma:\:0.82,\ \delta\!A_Z:     \:0.22,\
          \delta\!B_\gamma:\:0.10,\ \delta\!B_Z:     \:0.65 \ }_{\ 
          \delta  C_\gamma:\:0.11,\ \delta\!D_\gamma:\:1.76,\
          \delta\!D_Z:     \:14.6,\:\phantom{\delta}f_2^R:\:0.03 \
                   }\Bigr),} \non\\
&&{\mit\Delta}[\:\re(\delta\!D_\gamma)\:]=0.08\ \ \ \ \ \ \ 
  {\rm for}\ P_{e^-}=0.2\ {\rm and}\ P_{e^+}=0.1, \non\\
&&{\ \ \ \
\Bigl(^{\ \delta\!A_\gamma:\:0.52,\ \delta\!A_Z:     \:0.13,\
          \delta\!B_\gamma:\:0.09,\ \delta\!B_Z:     \:0.42 \ }_{\ 
          \delta  C_\gamma:\:0.15,\ \delta  C_Z:     \:1.13,\
          \delta\!D_Z:     \:14.4,\:\phantom{\delta}f_2^R:\:0.03 \
                   }\Bigr),} \non\\
&&{\mit\Delta}[\:\re(\delta\!D_Z)\:]=14.4\ \ \ \ \ \ \ 
  {\rm for}\ P_{e^-}=0.2\ {\rm and}\ P_{e^+}=0.1, \non\\
&&{\ \ \ \
\Bigl(^{\ \delta\!A_\gamma:\:0.52,\ \delta\!A_Z:     \:0.13,\
          \delta\!B_\gamma:\:0.09,\ \delta\!B_Z:     \:0.42 \ }_{\ 
          \delta  C_\gamma:\:0.15,\ \delta  C_Z:     \:1.13,\
          \delta\!D_\gamma:\:0.08,\:\phantom{\delta}f_2^R:\:0.03 \
                   }\Bigr),} 
\end{eqnarray}
where we also showed the expected precision of the other parameter
measurements for the same beam polarizations. For instance, we can
expect ${\mit\Delta}[\:\re(\delta\!A_\gamma)\:]=0.16$ for $P_{e^-}
=P_{e^+}=0.7$ while the expected precision of $\re(\delta\!A_Z)$,
$\re(\delta\!B_\gamma)$, $\cdots$ for the same polarizations are 0.13,
0.25, $\cdots$, respectively. This result is independent of the choice
of the non-SM parameters in contrast to the preceding results.

As it is seen the precision of $\delta\{C,D\}_Z$ measurement would be 
very poor even for the optimal polarization. This is mainly a
consequence of the size of ${\cal F}^{\ssl(*)}_{\{C,D\}Z}$, which is
illustrated in figs.\ref{co-func00} and \ref{co-func55}: These two
functions are very small in a large area. More quantitatively, the
size of the elements of $M$ matrix, $M_{ij}$, is $O(1)$ for $i,j \neq
7,9$, while the size of $M_{i7}(=M_{7i})$ and $M_{i9}(=M_{9i})$ is at
most $O(10^{-2})$. In addition, determination of $\delta\!D_\gamma$
would be practically difficult, as well, since its error varies
rapidly with the polarization. For example, ${\mit\Delta}[\:\re(\delta
\!D_\gamma)\:]$ becomes 0.86 for $P_{e^-}=0.1/P_{e^+}=0.1$ and 0.99
for $P_{e^-}=0.3/P_{e^+}=0.1$. The source of that sensitivity is
hidden in the neutral-current structure with $\sin^2\theta_W\simeq
0.23$. Indeed, the optimal polarization becomes $P_{e^-}=0.1$ instead
of 0.2 $({\mit\Delta} [\:\re(\delta\!D_\gamma)\:]=0.09)$ for $\sin^2
\theta_W=0.25$. On the other hand, a good determination (almost
independently of the polarization) could be expected for $f_2^R$.
Indeed, the best precision is
\begin{equation}
{\mit\Delta}[\:\re(f_2^R)\:]=0.01\ \ \ \ \ \ \ 
  {\rm for}\ P_{e^-}=-0.8\ {\rm and}\ P_{e^+}=-0.8
\end{equation}
whereas even for the unpolarized beams we obtain
${\mit\Delta}[\:\re(f_2^R)\:]=0.03$.

At the time a linear collider will be operating, data from Tevatron
Run II and LHC will also provide independent constraints on top-quark
couplings. Below we provide an example of a combined analysis assuming 
$\delta\!A_v$, $\delta\!B_v$ and $f_2^R$ are known and OO are used
to determine $\delta C_v$ and $\delta\!D_v$ only (here we put $\delta
A_v=\delta\!B_v=f_2^R=0$ for simplicity). The results are as follows:
\begin{equation}
\begin{array}{ll}
{\mit\Delta}[\:\re(\delta C_\gamma)\:]=0.04\ \ \ \ \ \ \ &
  {\rm for}\ P_{e^-}=0.2\ {\rm and}\ P_{e^+}=0.2 \\
{\mit\Delta}[\:\re(\delta C_Z)\:]=0.23 &
  {\rm for}\ P_{e^-}=0.2\ {\rm and}\ P_{e^+}=0.1 \\
{\mit\Delta}[\:\re(\delta\!D_\gamma)\:]=0.03\ \ \ \ \ \ \ &
  {\rm for}\ P_{e^-}=0.2\ {\rm and}\ P_{e^+}=0.1 \\
{\mit\Delta}[\:\re(\delta\!D_Z)\:]=2.97 &
  {\rm for}\ P_{e^-}=0.2\ {\rm and}\ P_{e^+}=0.1 \\
\end{array}
\end{equation}
The error for $\delta\!D_Z$ became much smaller but still too large
for practical use. However, as we have seen in sec.3, the
$C\!P$-sensitive asymmetry ${\cal A}_{\sst{C\!P}}^{\ssf}$ would
provide much stronger constraints on $\delta\!D_{\gamma,Z}$.

\sec{Summary and Conclusions}

We have presented here the angular and energy distributions for $\fb$
in the process $\epem \to \ttbar \to \fb \cdots$, where $f=\ell$ or
$b$ quark in the form suitable for an application of the optimal
observables (OO). The most general ($C\!P$-violating {\it and}
$C\!P$-conserving) couplings for $\gamma\ttbar$, $Z \ttbar$ and $Wtb$
have been assumed. All fermion masses except $m_t$ have been neglected
and we have kept only terms linear in anomalous couplings. We have
assumed the tagging efficiency at the level of $60\%$ both for lepton
and $b$ quark detection, the range of the polar angle restricted by
$|\cos\thf|<0.9$ and the integrated luminosity $L=500 \fbarn^{-1}$.

$C\!P$-violating charge forward-backward asymmetry 
${\cal A}_{\sst{C\!P}}^{\ssf}$ has been introduced as an efficient way
for testing $C\!P$-violation in top-quark couplings. Since the angular
distribution for leptons is insensitive to variations of the standard
V$-$A structure of the $Wtb$ coupling, the asymmetry could be utilized
for a pure test of $C\!P$-violation in the top-quark production
process. The expected statistical significance  $N_{\!\sst{S\!D}}$ for
the measurement of the asymmetry has been calculated. We have found
that it should be possible to detect ${\cal A}_{\sst{C\!P}}^{\ssf}$ at
$5.5\sigma$ ($4.3\sigma$) level for bottom quarks (leptons) for
unpolarized beams, assuming $C\!P$-violating couplings of the order of
$0.05$. Having both beams polarized at $80\%$ the signal for bottom
quarks (leptons) could reach even $16\sigma$ ($6.6\sigma$).

Next, the OO procedure has been applied to the angular distributions.
In the case of the lepton angular distribution, the expected
statistical significance for signals of non-standard physics varies
between $1.3\sigma$ and 35.5$\sigma$ assuming non-standard form
factors of the order of 0.05. It turned out that in the case of the
bottom-quark angular distribution the statistical significance of the
signal is in general higher than for leptons because of larger event
rate and varies between $1.8\sigma$ and 85.1$\sigma$ for the same
non-standard form factors.

When deriving the above results we have fixed all the non-SM parameters
to be $+0.05$ as a reasonable example for the strength of beyond-the-SM
physics. However, final results for statistical significances
considered here depend on the size of the non-standard parameters. The
most convenient beam polarizations for a measurement of the asymmetry
${\cal A}_{\sst{C\!P}}^{\ssf}$ and for testing the angular
distributions  varies with the non-standard parameters, as well. 
Therefore one should stress that the beam polarizations should be 
carefully adjusted for each model to be tested in actual experimental 
analysis. However, in any case, the above results show that a
measurement of ${\cal A}^{\ssf}_{\sst{C\!P}}$ and OO analysis of the
angular distributions are both very efficient for  new-physics search.

Then we have analyzed the angular and energy distribution of the
lepton toward separate determinations of the anomalous form factors.
In order to reach the highest precision we have been adjusting beam
polarizations to minimize errors for each form factor. We have found
that at $\sqrt{s}=500$ GeV with the integrated luminosity $L=500
\fbarn^{-1}$ the best determined coupling would be the axial coupling
of the $Z$ boson with the error ${\mit\Delta}[\:\re(\delta\!A_Z)\:]=
0.07$ while the lowest precision is expected for $\re(\delta\!D_Z)$
with ${\mit\Delta}[\:\re(\delta\!D_Z)\:]=14.4$. This result is
independent of the choice of the non-SM parameters in contrast to the
above two types of analyses.

Concluding, we have observed that the angular distributions and the
angular and energy distributions of top-quark decay products both
provide very efficient tools for studying top-quark couplings to gauge
bosons at linear colliders.

\vspace*{0.6cm}
\centerline{ACKNOWLEDGMENTS}

\vspace*{0.3cm}
We would like to thank K.~Fujii for useful discussions concerning
details of lepton and bottom-quark detection and J.~Pliszka for his
remarks on the statistical analysis. One of us (Z.H.) is grateful to
Y.~Sumino, T. Nagano, T. Takahashi and K. Ikematsu for stimulating
discussions. This work is supported in part by the State Committee for
Scientific Research (Poland) under grant 2~P03B~014~14 and by Maria
Sk\l odowska-Curie Joint Fund II (Poland-USA) under grant
MEN/NSF-96-252. 

\vspace*{0.8cm}

\vspace*{0.8cm}
\renewcommand{\theequation}{A.\arabic{equation}}
\setcounter{equation}{0}
\noindent \hspace*{-0.72cm}
{\bf Appendix}

\noindent
Integrals of ${\mit\Theta}^{\ssf(*)}_i(x)$ denoted in the main text by
${\mit\Omega}^{\ssf(*)}_i$ in the angular distribution eq.(\ref{dis2})
are the following:
\begin{eqnarray}
&&{\mit\Omega}^{\ssf(*)}_0=
D_V^{(*)}-(1-2\beta^2)D_A^{(*)}-2\:{\rm Re}(G_1^{(*)}) \non\\
&&\phantom{{\mit\Omega}^{\ssf(*)}_0}
-\alpha^{\ssf}[\:2(1-\beta^2){\rm Re}(D_{V\!\!A}^{(*)}) 
-{\rm Re}(F_1^{(*)})+(3-2\beta^2){\rm Re}(G_3^{(*)})\:] \non\\
&&\phantom{{\mit\Omega}^{\ssf(*)}_0}
+\Bigl[\:D_V^{(*)}+D_A^{(*)}+2\:{\rm Re}(G_1^{(*)})  \non\\
&&\phantom{{\mit\Omega}^{\ssf(*)}_0}\ \ \ \ \ \ \ \ \
+\alpha^{\ssf}{\rm Re}(2D_{V\!\!A}^{(*)}-F_1^{(*)}+3G_3^{(*)})\:\Bigr]
\frac{1-\beta^2}{2\beta}\ln\frac{1+\beta}{1-\beta},  \non\\
&&{\mit\Omega}^{\ssf(*)}_1=4\:{\rm Re}(E_{V\!\!A}^{(*)})
+2\alpha^{\ssf}[\:(1-\beta^2)E_A^{(*)}
-{\rm Re}(F_4^{(*)}-G_2^{(*)})\:]
\non\\
&&\phantom{{\mit\Omega}^{\ssf(*)}_0}
-\bigl\{2{\rm Re}(E_{V\!\!A}^{(*)})+
\alpha^{\ssf}[\:E_V^{(*)}+E_A^{(*)}-{\rm Re}(F_4^{(*)}-G_2^{(*)})\:]
\:\bigr\}
\frac{1-\beta^2}{\beta}\ln\frac{1+\beta}{1-\beta}, \non\\
&&{\mit\Omega}^{\ssf(*)}_2=(3-2\beta^2)[\:D_V^{(*)}+D_A^{(*)}+
2{\rm Re}(G_1^{(*)})\:] \non\\
&&\phantom{{\mit\Omega}^{\ssf(*)}_2}
+3\alpha^{\ssf}[\:2(1-\beta^2)
{\rm Re}(D_{V\!\!A}^{(*)})-{\rm Re}(F_1^{(*)})
+(3-2\beta^2){\rm Re}(G_3^{(*)})\:] \non\\
&&\phantom{{\mit\Omega}^{\ssf(*)}_2}
-3\Bigl[\:D_V^{(*)}+D_A^{(*)}+2\:{\rm Re}(G_1^{(*)}) \non\\
&&\phantom{{\mit\Omega}^{\ssf(*)}_0}\ \ \ \ \ \ \ \ \
+\alpha^{\ssf}{\rm Re}(2D_{V\!\!A}^{(*)}-F_1^{(*)}+3G_3^{(*)})\:\Bigr]
\frac{1-\beta^2}{2\beta}\ln\frac{1+\beta}{1-\beta}.
\label{omegas}
\end{eqnarray}

Next we present explicit formulas of the coefficient functions
for the nine anomalous form factors in eq.(\ref{reexp})
${\cal F}_{\{A,B,C,D\}\{\gamma,Z\}}^{\ssf(*)}(x, \theta)$ and
${\cal F}_{2R}^{\ssf(*)}(x, \theta)$ ($f=\ell/b$):
\begin{eqnarray}
&&{\cal F}_{Av}^{\ssf(*)}(x, \theta) \non\\
&&=\Bigl[\:\frac12(3-\beta^2)C(D_V\!:\!A_v)f^{\ssf}(x)
  +2\alpha_0^{\ssf} C(D_{V\!\!A}\!:\!A_v)g^{\ssf}(x)\:\Bigr]
  (1+\cos^2\theta) \non\\
&&-\Bigl[\:\alpha_0^{\ssf}(1-\beta^2)C(D_{V\!\!A}\!:\!A_v)f^{\ssf}(x)
\non\\
&&\ \ \ \ \ \ \
  -\frac12\{C(D_V\!:\!A_v)+2\alpha_0^{\ssf} C(D_{V\!\!A}\!:\!A_v)\}
  \{2h_1^{\ssf}(x)-h_2^{\ssf}(x)\}\:\Bigr](1-3\cos^2\theta) \non\\
&&+2\Bigl[\:\alpha_0^{\ssf} C(E_V\!:\!A_v) \{g^{\ssf}(x)-h_1^{\ssf}(x) \}
  +2C(E_{V\!\!A}\!:\!A_v)\{ f^{\ssf}(x)-h_1^{\ssf}(x) \}\:\Bigr]
  \cos\theta, \non\\
&& \\
&&{\cal F}_{Bv}^{\ssf(*)}(x, \theta) \non\\
&&=\frac12\beta^2 C(D_{\!A}\!:\!B_v)f^{\ssf}(x)(3-\cos^2\theta)
  +2\alpha_0^{\ssf} C(D_{V\!\!A}\!:\!B_v)g^{\ssf}(x)(1+\cos^2\theta)
\non\\
&&-\frac12\Bigl[\:\{C(D_{\!A}\!:\!B_v)
  +2\alpha_0^{\ssf}(1-\beta^2)C(D_{V\!\!A}\!:\!A_v)\}f^{\ssf}(x)
\non\\
&&\ \ \ \ \ \ \
  -\{C(D_{\!A}\!:\!B_v)+2\alpha_0^{\ssf} C(D_{V\!\!A}\!:\!B_v)\}
   \{2h_1^{\ssf}(x)-h_2^{\ssf}(x)\}\:\Bigr]
  (1-3\cos^2\theta) \non\\
&&+2\Bigl[\:\{\alpha_0^{\ssf}(1-\beta^2)C(E_{\!A}\!:\!B_v)
  +2C(E_{V\!\!A}\!:\!B_v)\}f^{\ssf}(x)
  +\alpha_0^{\ssf} C(E_{\!A}\!:\!B_v)g^{\ssf}(x) \non\\
&&\ \ \ \ \ \ \
  -\{\alpha_0^{\ssf} C(E_{\!A}\!:\!B_v)
  +2C(E_{V\!\!A}\!:\!B_v)\}h_1^{\ssf}(x)\:\Bigr]\cos\theta, \\
&& \non\\
&&{\cal F}_{Cv}^{\ssf(*)}(x, \theta) \non\\
&&=-\beta^2 C(G_1\!:\!C_v)f^{\ssf}(x)(1+\cos^2\theta) \non\\
&&+2\alpha_0^{\ssf} C(G_2\!:\!C_v)
  \Bigl[\:f^{\ssf}(x)+g^{\ssf}(x)-h_1^{\ssf}(x)\:\Bigr]\cos\theta
\non\\
&&-\Bigl[\:\{C(G_1\!:\!C_v)+\alpha_0^{\ssf}(2-\beta^2)C(G_3\!:\!C_v)\}
  f^{\ssf}(x)+\alpha_0^{\ssf} C(G_3\!:\!C_v)g^{\ssf}(x) \non\\
&&\ \ \ \ \ \ \
  -\{2C(G_1\!:\!C_v)+3\alpha_0^{\ssf} C(G_3\!:\!C_v)\}h_1^{\ssf}(x)
\non\\
&&\ \ \ \ \ \ \
  +\{C(G_1\!:\!C_v)
  +\alpha_0^{\ssf} C(G_3\!:\!C_v)\}h_2^{\ssf}(x)\:\Bigr]
  (1-3\cos^2\theta) \\
&& \non\\
&&{\cal F}_{Dv}^{\ssf(*)}(x, \theta) \non\\
&&=\alpha_0^{\ssf}C(F_1\!:\!D_v)
  \Bigl[\:f^{\ssf}(x)-h_1^{\ssf}(x)\:\Bigr](1-3\cos^2\theta)
  -\alpha_0^{\ssf} C(F_1\!:\!D_v)g^{\ssf}(x)(1+\cos^2\theta) \non\\
&&\ \ \ \ \ \ \
  -2\alpha_0^{\ssf} C(F_4\!:\!D_v)\Bigl[\:f^{\ssf}(x)+g^{\ssf}(x)
  -h_1^{\ssf}(x)\:\Bigr]\cos\theta,
\end{eqnarray}
while ${\cal F}_{2R}^{\ell(*)}(x, \theta)$ takes different forms for
$f=\ell$ and $f=b$ as
\begin{eqnarray}
&&{\cal F}_{2R}^{\ell(*)}(x, \theta)\non\\
&&=\frac12\Bigl[\:(3-\beta^2)D^{(0,*)}_V -(1-3\beta^2)D^{(0,*)}_{\!A}
  -2(1-\beta^2){\rm Re}(D^{(0,*)}_{V\!\!A})\:\Bigr]
  \:\delta\!f^{\ssl}(x) \non\\
&&+2{\rm Re}(D^{(0,*)}_{V\!\!A})\:\delta g^{\ssl}(x)
  (1+\cos^2\theta) \non\\
&&+\frac12\Bigl[\:D^{(0,*)}_V+D^{(0,*)}_{\!A}
  +2{\rm Re}(D^{(0,*)}_{V\!\!A})\:\Bigr]
  \Bigl[\:2\delta h_1^{\ssl}(x)-\delta h_2^{\ssl}(x)\:\Bigr]
  (1-3\cos^2\theta) \non\\
&&+2\Bigl[\:(1-\beta^2)E^{(0,*)}_{\!A}
+2{\rm Re}(E^{(0,*)}_{V\!\!A})\:\Bigr]\delta\!f^{\ssl}(x)\cos\theta
\non\\
&&+2\:(\:E^{(0,*)}_V+E^{(0,*)}_{\!A}\:)\:
  \delta g^{\ssl}(x)\cos\theta \non\\
&&-2\:\Bigl[\:E^{(0,*)}_V
  +E^{(0,*)}_{\!A}+2{\rm Re}(E^{(0,*)}_{V\!\!A})
\:\Bigr]\:\delta h_1^{\ssl}(x)\cos\theta \non\\
&&+\frac12\Bigl[\:(3-\beta^2)(D^{(0,*)}_V +D^{(0,*)}_{\!A})
   +6(1-\beta^2){\rm Re}(D^{(0,*)}_{V\!\!A})\:\Bigr]\:
   \delta\!f^{\ssl}(x)\cos^2\theta,
\end{eqnarray}
and 
\begin{eqnarray}
&&{\cal F}_{2R}^{b(*)}(x, \theta) \non\\
&&=\alpha_1^{\ssb}\left\{{\rm Re}(D_{V\!\!A}^{(0,*)})
  \Bigl[\:-\Bigl\{(1-\beta^2)f^{\ssb}(x)-2h_1^{\ssb}(x)+h_2^{\ssb}(x)
  \Bigr\}(1-3\cos^2\theta)\right. \non\\
&&\phantom{=\alpha_1^{\ssb}{\rm Re}(D_{V\!\!A}^{(0,*)})\Bigl[\:}
  +2g^{\ssb}(x)(1+\cos^2\theta)\:\Bigr] \non\\
&&\left. +2\Bigl[\:(1-\beta^2)E_A^{(0,*)}f^{\ssb}(x)
+(E_V^{(0,*)}+E_A^{(0,*)})
  \Bigl\{g^{\ssb}(x)-h_1^{\ssb}(x)\Bigr\}\:\Bigr]\cos\theta\right\},
\end{eqnarray}
where the functions $f^{\ssf}(x)$, $g^{\ssf}(x)$, $h_{1,2}^{\ssf}(x)$, 
$\delta\!f^{\ssf}(x)$, $\delta g^{\ssf}(x)$ and
$\delta h_{1,2}^{\ssf}(x)$ are defined as
\begin{eqnarray}
&&F^{\ssf}(x)=f^{\ssf}(x)+{\rm Re}(f_2^R)\delta\!f^{\ssf}(x), \non\\
&&G^{\ssf}(x)=g^{\ssf}(x)+{\rm Re}(f_2^R)\delta g^{\ssf}(x), \non\\
&&H_{1,2}^{\ssf}(x)
=h_{1,2}^{\ssf}(x)+{\rm Re}(f_2^R)\delta h_{1,2}^{\ssf}(x),
\end{eqnarray}
with $F^{\ssf}(x)$, $G^{\ssf}(x)$ and $H_{1,2}^{\ssf}(x)$ being given
as follows \cite{GH_00}
\begin{eqnarray}
&&F^{\ssf}(x)\equiv\frac{1}{B_{\ssf}}
\int d\omega\frac1{{{\mit\Gamma}}_t}
\frac{d^2{{\mit\Gamma}}_{\ssf}}{dx d\omega},\ \ \
G^{\ssf}(x)\equiv\frac{1}{B_{\ssf}}\int d\omega
\Bigl[\:1-x\frac{1+\bet}{1-\omega}\Bigr]
\frac1{{{\mit\Gamma}}_t}
\frac{d^2{{\mit\Gamma}}_{\ssf}}{dx d\omega}, \non\\
&&H_1^{\ssf}(x)\equiv\frac{1}{B_{\ssf}}\frac{1-\bet}{x}
\int d\omega (1-\omega)\frac1{{{\mit\Gamma}}_t}
\frac{d^2{{\mit\Gamma}}_{\ssf}}{dx d\omega}, \non\\
&&H_2^{\ssf}(x)\equiv\frac{1}{B_{\ssf}}\Bigl(\frac{1-\bet}{x}
\Bigr)^2
\int d\omega (1-\omega)^2\frac1{{{\mit\Gamma}}_t}
\frac{d^2{{\mit\Gamma}}_{\ssf}}{dx d\omega},
\end{eqnarray}
and $\omega$ is defined as $\omega\equiv (p_t -p_{\ssf})^2/m^2_t$.

After performing the above integrations using
$$
\frac{1}{{\mit\Gamma}_t}\frac{d^2{\mit\Gamma}_{\ssf}}{dx d\omega}
=\left\{
\begin{array}{ll}
{\dps \frac{1+\beta}{\beta}\;\frac{3 B_{\ell}}{W}                         
\omega\left[\:1+2{\rm Re}(f_2^R)\sqrt{r}\left(\frac{1}{1-\omega}-      
\frac{3}{1+2r} \right)\:\right]}
 &  {\rm for}\;\;f={\ell}^+,  \\
 &                         \\
{\dps \frac{1+\beta}{2\beta(1-r)}\delta(\omega-r)}
 & 
{\rm for}\;\;f=b.
\end{array}
\right.
$$
one obtains the following explicit forms of $f^{\ssf}(x)$,
$g^{\ssf}(x)$, $h_{1,2}^{\ssf}(x)$, $\delta\!f^{\ssf}(x)$,
$\delta g^{\ssf}(x)$ and $\delta h_{1,2}^{\ssf}(x)$ for leptonic and
bottom-quark final states:\\
$\bullet$ For $f=\ell$
\begin{eqnarray*}
&&f^{\ssl}(x)
=\frac{3(1+\beta)}{2\beta W}[\:\omega^2\:]^{\omega_+}_{\omega_-}\ 
\Bigl(\equiv \frac{3(1+\beta)}{2\beta W}(\omega_+^2-\omega_-^2)\Bigr),
  \\
&&g^{\ssl}(x)=f^{\ssl}(x)+\frac{3(1+\beta)^2}{\beta W}x
\Bigl[\:\omega+\ln|1-\omega|\:\Bigr]^{\omega_+}_{\omega_-}, \\
&&h_1^{\ssl}(x)=\frac{1-\beta^2}{2\beta W}\frac1{x}
\Bigl[\:\omega^2 (3-2\omega)\:\Bigr]^{\omega_+}_{\omega_-}, \\
&&h_2^{\ssl}(x)=\frac1{4\beta W}(1+\beta)(1-\beta)^2\frac1{x^2}
  \Bigl[\:\omega^2(6-8\omega+3\omega^2)\:\Bigr]^{\omega_+}_{\omega_-},
\end{eqnarray*}
\begin{eqnarray}
&&\delta\!f^{\ssl}(x)=-\frac{3(1+\beta)}{\beta W}\sqrt{r}\Bigl[\:
2\omega+2\ln|1-\omega|+\frac{3\omega^2}{1+2r}\:
\Bigr]^{\omega_+}_{\omega_-}, \non\\
&&\delta g^{\ssl}(x)
  =\delta\!f^{\ssl}(x)-\frac{6(1+\beta)^2}{\beta W}\sqrt{r}\:x
  \Bigl[\:\ln|1-\omega| \non\\
&&\phantom{\delta g^{\ssl}(x)
=\delta\!f^{\ssl}(x)-\frac{6(1+\beta)^2}{\beta W}x\sqrt{r}\Bigl[}
+\frac1{1-\omega}
+\frac3{1+2r}(\omega+\ln|1-\omega|)\:\Bigr]^{\omega_+}_{\omega_-},
\non\\
&&\delta h_1^{\ssl}(x)=\frac{3(1-\beta^2)}{\beta W}\frac{\sqrt{r}}{x}
\Bigl[\:\omega^2
\Bigl(1-\frac{3-2\omega}{1+2r}\Bigr)\:\Bigr]^{\omega_+}_{\omega_-},
\non\\
&&\delta h_2^{\ssl}(x)=\frac1{2\beta W}(1+\beta)(1-\beta)^2  \non\\
&&\phantom{\delta h_2^{\ssl}(x)===}
\times\frac{\sqrt{r}}{x^2}\Bigl[\:2\omega^2(3-2\omega)
  -\frac{3\omega^2}{1+2r}(6-8\omega+3\omega^2)
  \:\Bigr]^{\omega_+}_{\omega_-},~~
\end{eqnarray}
where $\omega_{\pm}$ are given as follows: \\
For $r \geq B$\ \ ($r\equiv M_W^2/m_t^2$ and $B\equiv
(1-\beta)/(1+\beta)$)
\begin{equation}
\begin{array}{llll}
\omega_+=1-r, & \omega_-=1-x/B~~~~ 
                               & {\rm for} & Br\leq x < B   \\
\omega_+=1-r, & \omega_-=0     & {\rm for} & B \leq x < r   \\
\omega_+=1-x, & \omega_-=0     & {\rm for} & r\leq x \leq 1 \\
\end{array}
\end{equation}
For $r < B$
\begin{equation}
\begin{array}{llll}
\omega_+=1-r, & \omega_-=1-x/B~~~~ 
                               & {\rm for} & Br \leq x < r  \\
\omega_+=1-x, & \omega_-=1-x/B & {\rm for} & r \leq x < B   \\
\omega_+=1-x, & \omega_-=0     & {\rm for} & B \leq x \leq 1\\
\end{array}
\end{equation}
$\bullet$ For $f=b$
\begin{eqnarray}
&&f^{\ssb}(x)=\frac{1+\beta}{2\beta(1-r)}\:(={\rm constant}), \non\\
&&g^{\ssb}(x)
=\Bigl(1-\frac{1+\beta}{1-r}x\Bigr)\frac{1+\beta}{2\beta(1-r)},
\non\\
&&h_1^{\ssb}(x)=\frac{1-\beta^2}{2\beta x}, \non\\
&&h_2^{\ssb}(x)=\frac{(1-r)(1+\beta)(1-\beta)^2}{2\beta x^2}, \non\\
&&\delta\!f^{\ssb}(x)=\delta g^{\ssb}(x)=\delta h_1^{\ssb}(x)
=\delta h_2^{\ssb}(x)=0,
\end{eqnarray}
where $x$ is bounded as
$$
B(1-r) \leq x \leq 1-r.
$$

The coefficients $C(X:Y)$ employed in the definition of the
coefficient functions have been introduced through the following
formulas:
\begin{eqnarray}
&&D_V^{(*)}=D_V^{(0,*)}
+\sum_{v=\gamma,Z}C(D_V\!:\!A_v){\rm Re}(\delta\!A_v), \non\\
&&D_{\!A}^{(*)}=D_{\!A}^{(0,*)}
+\sum_{v=\gamma,Z}C(D_{\!A}\!:\!B_v){\rm Re}(\delta\!B_v), \non\\
&&{\rm Re}(D_{V\!\!A}^{(*)})={\rm Re}(D_{V\!\!A}^{(0,*)}) \non\\
&&\phantom{{\rm Re}(D_{V\!\!A}^{(*)})=}
+\sum_{v=\gamma,Z}\Bigl[\:
 C(D_{V\!\!A}\!:\!A_v){\rm Re}(\delta\!A_v)
+C(D_{V\!\!A}\!:\!B_v){\rm Re}(\delta\!B_v)\:\Bigr],
\end{eqnarray}
and in the analogous manner for $E_{V,A,V\!\!A}$, $F_{1\sim 4}$ and
$G_{1\sim 4}$. $D_{V,A,V\!\!A}^{(0,*)}$, $E_{V,A,V\!\!A}^{(0,*)}$,
$F_{1\sim 4}^{(0,*)}$, $G_{1\sim 4}^{(0,*)}$ could be obtained from
eq.(\ref{pol_coeff}) below as a SM approximation of
$D_{V,A,V\!\!A}^{(*)}$, $E_{V,A,V\!\!A}^{(*)}$, $F_{1\sim 4}^{(*)}$,
$G_{1\sim 4}^{(*)}$. Explicit forms of the independent coefficients
are given as
\begin{eqnarray}
&&C(D_V\!:\!A_\gamma)
=2C[\:{\cal P}_{\otimes}A_\gamma
  -({\cal P}_{\oplus}+v_e{\cal P}_{\otimes})d'A_Z\:], \non\\
&&C(E_V\!:\!A_\gamma)
=-2C[\:{\cal P}_{\oplus}A_\gamma
  -({\cal P}_{\otimes}+v_e{\cal P}_{\oplus})d'A_Z\:], \non\\
&&C(D_{V\!\!A}\!:\!A_\gamma)
  =-C({\cal P}_{\oplus}+v_e{\cal P}_{\otimes})d'B_Z, \non\\
&&C(E_{V\!\!A}\!:\!A_\gamma)
  =C({\cal P}_{\otimes}+v_e{\cal P}_{\oplus})d'B_Z, \non\\
&&C(D_V\!:\!A_Z)
  =-2C[\:({\cal P}_{\oplus}+v_e{\cal P}_{\otimes})d'A_\gamma
  -\{2v_e{\cal P}_{\oplus}+(1+v_e^2){\cal P}_{\otimes}\}d'^2A_Z\:],
  \non\\
&&C(E_V\!:\!A_Z)
  =2C[\:({\cal P}_{\otimes}+v_e{\cal P}_{\oplus})d'A_\gamma
  -\{2v_e{\cal P}_{\otimes}+(1+v_e^2){\cal P}_{\oplus}\}d'^2A_Z\:],
  \non\\
&&C(D_{V\!\!A}\!:\!A_Z)=C[\:2v_e{\cal P}_{\oplus}
  +(1+v_e^2){\cal P}_{\otimes}\:]d'^2B_Z, \non\\
&&C(E_{V\!\!A}\!:\!A_Z)=-C[\:2v_e{\cal P}_{\otimes}
  +(1+v_e^2){\cal P}_{\oplus}\:]d'^2B_Z,
\label{DVA}
\end{eqnarray}
where $v_e=-1+4\sin^2\theta_W$, $d'\equiv s/[4\sin\theta_W
\cos\theta_W(s-M_Z^2)]$, two polarization factors ${\cal P}_{\oplus}$
and ${\cal P}_{\otimes}$ are defined as
\[
{\cal P}_{\oplus}\equiv P_{e^-}+P_{e^+},\ \ \ \ \ {\cal P}_{\otimes}
\equiv 1+P_{e^-}P_{e^+},
\]
and the others are thereby given as
\begin{equation}
\begin{array}{ll}
C(D_{\!A}\!:\!B_v)=2C(D_{V\!\!A}\!:\!A_v),\ \ \ \ \
&  C(E_{\!A}\!:\!B_v)=2C(E_{V\!\!A}\!:\!A_v),\\
C(D_{V\!\!A}\!:\!B_v)=C(D_V\!:\!A_v)/2,
&  C(E_{V\!\!A}\!:\!B_v)=C(E_V\!:\!A_v)/2,\\
C(G_1\!:\!C_v)=C(D_V\!:\!A_v)/2,
&  C(G_2\!:\!C_v)=C(E_V\!:\!A_v)/2,\\
C(G_3\!:\!C_v)=C(D_{V\!\!A}\!:\!A_v),
&  C(G_4\!:\!C_v)=C(E_{V\!\!A}\!:\!A_v),\\
C(F_1\!:\!D_v)=-C(D_V\!:\!A_v)/2,
&  C(F_2\!:\!D_v)=-C(E_V\!:\!A_v)/2,\\
C(F_3\!:\!D_v)=-C(D_{V\!\!A}\!:\!A_v),
&  C(F_4\!:\!D_v)=-C(E_{V\!\!A}\!:\!A_v).\\
\end{array}
\label{other-coeff}
\end{equation}

As explained in the main text, they are not always independent
of each other. When $P_{\oplus}=\pm P_{\otimes}$, i.e., $P_{e^-}
=P_{e^+}=\pm 1$, we have 
\[
C(\{D_V,E_V,D_{V\!\!A},E_{V\!\!A}\}\!:\!A_Z)
=\mp (1\pm v_e)d'C(\{D_V,E_V,D_{V\!\!A},E_{V\!\!A}\}\!:\!A_\gamma),
\]
As a consequence of the above relations one gets
\[
{\cal F}_{\{A,B,C,D\}Z}^{\ssf(*)}(x, \theta)
=\mp (1\pm v_e)d'{\cal F}_{\{A,B,C,D\}\gamma}^{\ssf(*)}(x, \theta).
\]
In this case all we can determine (for the production form factors)
are the following four combinations
\[
{\rm Re}(\delta\{A,B,C,D\}_\gamma \mp
(1 \pm v_e)d'\delta\{A,B,C,D\}_Z).
\]

Finally we present here formulas for 
$D_{V,A,V\!\!A}^{(*)}$, $E_{V,A,V\!\!A}^{(*)}$, $F_{1\sim 4}^{(*)}$, 
$G_{1\sim 4}^{(*)}$ for completeness:
\begin{eqnarray}
&&D_{V,\,A,\,V\!\!A}^{(*)}={\cal P}_{\otimes}D_{V,\,A,\,V\!\!A}
-{\cal P}_{\oplus}E_{V,\,A,\,V\!\!A}, \non\\
&&E_{V,\,A,\,V\!\!A}^{(*)}={\cal P}_{\otimes}E_{V,\,A,\,V\!\!A}
-{\cal P}_{\oplus}D_{V,\,A,\,V\!\!A}, \non\\
&&F_{1,\,2,\,3,\,4}^{(*)}= {\cal P}_{\otimes}F_{1,\,2,\,3,\,4}
-{\cal P}_{\oplus}F_{2,\,1,\,4,\,3}, \non\\
&&G_{1,\,2,\,3,\,4}^{(*)}= {\cal P}_{\otimes}G_{1,\,2,\,3,\,4}
-{\cal P}_{\oplus}G_{2,\,1,\,4,\,3},
\label{pol_coeff}
\end{eqnarray}
for
\begin{eqnarray}
&& D_V\:\equiv \:C\,[\,\avs-2\av\az\ci+\azs\cz+2(\av-\az\ci)
{\rm Re}(\delta\!A_\gamma)
\non\\
&& \lspace-2\{ \av\ci-\az\cz \}{\rm Re}(\delta\!A_Z)\,],
\non\\
&& D_A\:\equiv \:C\,[\,B_Z^2(1+v_e^2)\cci^2
-2B_Z v_e \cci{\rm Re}(\delta\!B_\gamma)
+2B_Z(1+v_e^2)\cci^2{\rm Re}(\delta\!B_Z)\,],
\non\\
&& D_{V\!\!A}\:\equiv \:C\,
[\,-\av\bz\ci+\az\bz\cz-\bz\ci(\delta\!A_\gamma)^*
\non\\
&& \lspace
+(A_\gamma -v_e \cci A_Z)\delta\!B_\gamma +\bz\cz(\delta\!A_Z)^* 
\non\\
&& \lspace-\{ \av\ci-\az\cz \}\delta\!B_Z \,],
\non\\
&& E_V\:\equiv \:2C\,
[\,\av\az\cci-\azs\ccz+\az\cci{\rm Re}(\delta\!A_\gamma)
+(\av\cci-2\az\ccz){\rm Re}(\delta\!A_Z) \,],
\non\\
&& E_A\:\equiv \:2C\,[\,-\bzs\ccz+B_Z \cci{\rm Re}(\delta\!B_\gamma)
-2\bz\ccz{\rm Re}(\delta\!B_Z) \,],
\non\\
&& E_{V\!\!A}\:\equiv \:C\,
[\,\av\bz\cci-2\az\bz\ccz+\bz\cci(\delta\!A_\gamma)^*
+A_Z \cci\delta\!B_\gamma
\non\\
&& \lspace-2\bz\ccz(\delta\!A_Z)^*
+(\av\cci-2\az\ccz)\delta\!B_Z \,],
\non\\
&& F_1\:\equiv \:C\,[\, -(\av-\az\ci)\delta\!D_\gamma
     +\{\av\ci-\az\cz\}\delta\!D_Z \,],
\non\\
&& F_2\:\equiv \:C\,[\, -\az\cci\delta\!D_\gamma
     -(\av\cci-2\az\ccz)\delta\!D_Z \,],
\non\\
&& F_3\:\equiv \:C\,[\, \bz\ci\delta\!D_\gamma-\bz\cz\delta\!D_Z \,],
\non\\
&& F_4\:\equiv \:C\,
[\, -\bz\cci\delta\!D_\gamma+2\bz\ccz\delta\!D_Z \,],
\non\\
&& G_1\:\equiv \:C\,[\, (\av-\az\ci)\delta C_\gamma
     -\{\av\ci-\az\cz\}\delta C_Z \,],
\non\\
&& G_2\:\equiv \:C\,[\, \az\cci\delta C_\gamma
     +(\av\cci-2\az\ccz)\delta C_Z \,],
\non\\
&& G_3\:\equiv \:C\,[\, -\bz\ci\delta C_\gamma +\bz\cz\delta C_Z \,],
\non\\
&& G_4\:\equiv \:C\,
[\, \bz\cci\delta C_\gamma-2\bz\ccz\delta C_Z \,]
\label{coefficients}
\end{eqnarray}
with $C \equiv 1/(4\sin^2\theta_W)$.

\vspace*{1.2cm}
\centerline{\large\bf $*\!*\!*\!*\!*$ Note added after Publication $*\!*\!*\!*\!*$}

\renewcommand{\theequation}{A.\arabic{equation}}
\setcounter{equation}{2}
\setcounter{figure}{0}

\vspace*{0.5cm} \noindent
{\bf [ Corrigendum ]} After this article has been published in \npb{585}{2000}{3},
we have found that equation (\ref{eqA3}) contains an error:\
$C(D_{V\!\!A}\!:\!A_v)$ in the third line should be replaced with
$C(D_{V\!\!A}\!:\!B_v)$ as
\begin{eqnarray}
&&{\cal F}_{Bv}^{\ssf(*)}(x, \theta) \non\\
&&=\frac12\beta^2 C(D_{\!A}\!:\!B_v)f^{\ssf}(x)(3-\cos^2\theta)
  +2\alpha_0^{\ssf} C(D_{V\!\!A}\!:\!B_v)g^{\ssf}(x)(1+\cos^2\theta)
\non\\
&&-\frac12\Bigl[\:\{C(D_{\!A}\!:\!B_v)
  +2\alpha_0^{\ssf}(1-\beta^2)C(D_{V\!\!A}\!:\!B_v)\}f^{\ssf}(x)
\non\\
&&\ \ \ \ \ \ \
  -\{C(D_{\!A}\!:\!B_v)+2\alpha_0^{\ssf} C(D_{V\!\!A}\!:\!B_v)\}
   \{2h_1^{\ssf}(x)-h_2^{\ssf}(x)\}\:\Bigr]
  (1-3\cos^2\theta) \non\\
&&+2\Bigl[\:\{\alpha_0^{\ssf}(1-\beta^2)C(E_{\!A}\!:\!B_v)
  +2C(E_{V\!\!A}\!:\!B_v)\}f^{\ssf}(x)
  +\alpha_0^{\ssf} C(E_{\!A}\!:\!B_v)g^{\ssf}(x) \non\\
&&\ \ \ \ \ \ \
  -\{\alpha_0^{\ssf} C(E_{\!A}\!:\!B_v)
  +2C(E_{V\!\!A}\!:\!B_v)\}h_1^{\ssf}(x)\:\Bigr]\cos\theta\,. \label{eqA3}
\end{eqnarray}

\renewcommand{\theequation}{4.\arabic{equation}}
\setcounter{equation}{12}

\noindent
Due to this correction, the two graphs expressing ${\cal F}_{B\gamma}^{\ssl(*)}$
and ${\cal F}_{BZ}^{\ssl(*)}$ in Figs.1 and 2 are to be replaced with those
presented below:

\vfill 

\begin{figure}[h]                                                    
\postscript{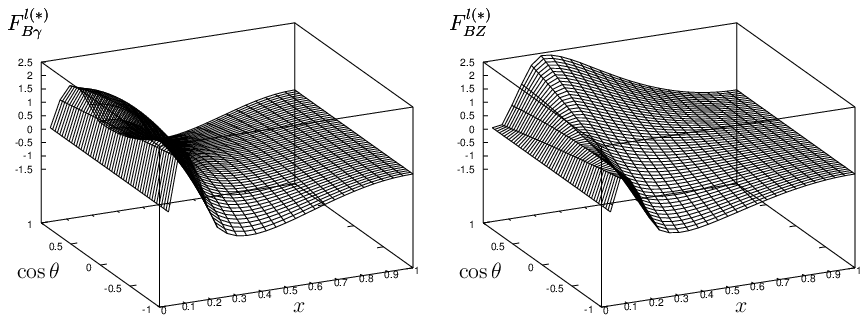}{0.75}
\vspace*{0.3cm}    
\caption{The shape of ${\cal F}^{\ssl(*)}_{B\{\gamma,Z\}}$
for unpolarized beams}
\end{figure}
%
\newpage 
%
\begin{figure}
\postscript{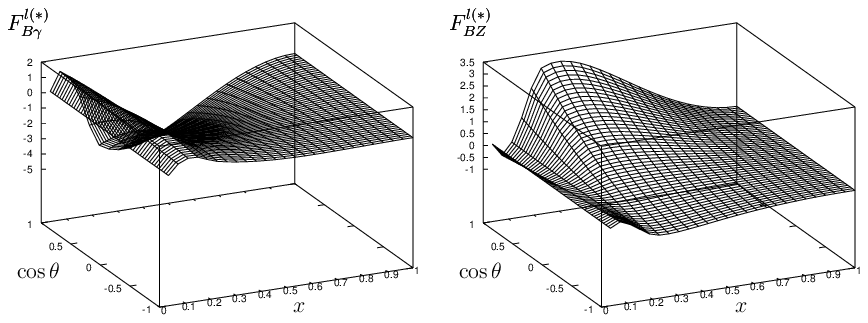}{0.75}
\vspace*{0.3cm}    
\caption{The shape of ${\cal F}^{\ssl(*)}_{B\{\gamma,Z\}}$
for $P_{e^-}=P_{e^+}=0.5$}
\end{figure}
 
\vspace*{-0.3cm}

The numerical results shown in Eqs.(\ref{eq413}) and (\ref{eq414}) are also
no longer valid, and we have carried out re-computations. Concerning the
former, i.e., Eq.(\ref{eq413}), after correcting the error we find very
large statistical uncertainties for measurements of the nine independent
non-SM parameters, therefore, in practice it will be impossible (with no
other experimental input) to determine all of them at once through the
distribution that was considered, i.e., the one in Eq.(\ref{reexp}).

Among those non-SM couplings, however, $\delta\!A_\gamma$ term is directly
related to the top-quark electric charge and expected to be studied in various
other ways. We therefore would like to give the results of an analysis without
$\delta\!A_\gamma$ term and replace Eq.(\ref{eq413}) with
\begin{eqnarray}
&&{\mit\Delta}[\:\re(\delta\!A_Z)\:]=4.0\times 10^{-2}\ \ \ \ \ \ \
  {\rm for}\ P_{e^-}/P_{e^+}=0.4/0.4, \non\\
&&{\ \ \ \
\Bigl(^{\ \ ******            \ \ \ \delta\!B_\gamma:\:0.08,\
          \delta\!B_Z:     \:0.06,\ \delta  C_\gamma:\:0.52 \ }_{\ 
          \delta  C_Z:     \:1.47,\ \delta\!D_\gamma:\:5.25,\
          \delta\!D_Z:     \:17.8,\:\phantom{\delta}f_2^R:\:0.02 \
                   }\Bigr),} \non\\
&&{\mit\Delta}[\:\re(\delta\!B_\gamma)\:]=7.2\times 10^{-2}\ \ \ \ \ \ \ 
  {\rm for}\ P_{e^-}/P_{e^+}=0.2/0.3,\ 0.3/0.2, \non\\
&&{\ \ \ \
\Bigl(^{\ \delta\!A_Z:     \:0.04,\ \ ******            \ \ \
          \delta\!B_Z:     \:0.05,\ \delta  C_\gamma:\:0.25 \ }_{\ 
          \delta  C_Z:     \:1.17,\ \delta\!D_\gamma:\:1.86,\
          \delta\!D_Z:     \:14.6,\:\phantom{\delta}f_2^R:\:0.03 \
                   }\Bigr),} \non\\
&&{\mit\Delta}[\:\re(\delta\!B_Z)\:]=4.5\times 10^{-2}\ \ \ \ \ \ \ 
  {\rm for}\ P_{e^-}/P_{e^+}=0.2/0.3,\ 0.3/0.2, \non\\
&&{\ \ \ \
\Bigl(^{\ \delta\!A_Z:     \:0.04,\ \delta\!B_\gamma:\:0.07,\
          \ ******            \ \ \ \delta  C_\gamma:\:0.25 \ }_{\ 
          \delta  C_Z:     \:1.17,\ \delta\!D_\gamma:\:1.86,\
          \delta\!D_Z:     \:14.6,\:\phantom{\delta}f_2^R:\:0.03 \
                   }\Bigr),} \non\\
&&{\mit\Delta}[\:\re(\delta\!C_\gamma)\:]=1.0\times 10^{-1}\ \ \ \ \ \ \ 
  {\rm for}\ P_{e^-}/P_{e^+}=0.1/0.1, \non\\
&&{\ \ \ \
\Bigl(^{\ \delta\!A_Z:     \:0.06,\ \delta\!B_\gamma:\:0.08,\
          \delta\!B_Z:     \:0.07,\ \ ******            \ \ \ }_{\ 
          \delta  C_Z:     \:1.07,\ \delta\!D_\gamma:\:0.81,\
          \delta\!D_Z:     \:13.9,\:\phantom{\delta}f_2^R:\:0.03 \
                   }\Bigr),} \non\\
&&{\mit\Delta}[\:\re(\delta C_Z)\:]=1.1\times 10^{0}\ \ \ \ \ \ \ \
  {\rm for}\ P_{e^-}/P_{e^+}=0.1/0.1, \non\\
&&{\ \ \ \
\Bigl(^{\ \delta\!A_Z:     \:0.06,\ \delta\!B_\gamma:\:0.08,\
          \delta\!B_Z:     \:0.07,\ \delta  C_\gamma:\:0.10 \ }_{\ 
          \ ******              \ \ \ \delta\!D_\gamma:\:0.81,\
          \delta\!D_Z:     \:13.9,\:\phantom{\delta}f_2^R:\:0.03 \
                   }\Bigr),} \non\\
&&{\mit\Delta}[\:\re(\delta D_\gamma)\:]=6.9\times 10^{-2}\ \ \ \ \ \ \ 
  {\rm for}\ P_{e^-}/P_{e^+}=0.1/0.2,\ 0.2/0.1, \non\\
&&{\ \ \ \
\Bigl(^{\ \delta\!A_Z:     \:0.05,\ \delta\!B_\gamma:\:0.07,\
          \delta\!B_Z:     \:0.06,\ \delta  C_\gamma:\:0.13 \ }_{\ 
          \delta  C_Z:     \:1.08,\ \ ******                 \ \ \
          \delta\!D_Z:     \:13.9,\:\phantom{\delta}f_2^R:\:0.03 \
                   }\Bigr),} \non\\
&&{\mit\Delta}[\:\re(\delta\!D_Z)\:]=1.4\times 10^{+1}\ \ \ \ \ \ \ 
  {\rm for}\ P_{e^-}/P_{e^+}=0.1/0.2,\ 0.2/0.1, \non\\
&&{\ \ \ \
\Bigl(^{\ \delta\!A_Z:     \:0.05,\ \delta\!B_\gamma:\:0.07,\
          \delta\!B_Z:     \:0.06,\ \delta  C_\gamma:\:0.13 \ }_{\ 
          \delta  C_Z:     \:1.08,\ \delta\!D_\gamma:\:0.07,\
          \ ******              \ \ \:\phantom{\delta}f_2^R:\:0.03 \
                   }\Bigr).} 
\label{eq413}
\end{eqnarray}
On the other hand, Eq.(\ref{eq414}) is simply to be replaced by
\begin{equation}
{\mit\Delta}[\:\re(f_2^R)\:]=1.5\times 10^{-2}\ \ \ \ \
  {\rm for}\ P_{e^-}=-0.9\ {\rm and}\ P_{e^+}=-0.9\,. \label{eq414}
\end{equation}

In spite of these modifications, conclusions concerning Eq.(\ref{eq413})
are not affected substantially and hold except for those on
$\delta\!A_\gamma$, if only we properly adjust the parameter values used
there according to the above corrected eqs.(\ref{eq413}) and (\ref{eq414}).

\vspace*{0.5cm}
We would like to thank very much Patrick Janot for kindly pointing out
that one term in eq.(\ref{eqA3}) seems unnatural and therefore might be
a typo. This led us to rechecking that equation and finding the error
mentioned here.

\vskip 1cm

\end{document}